%% file: BE2Helfrich.tex
\documentclass{ingosclass}
\bibliography{bib}

\newcommand{\clineX}[1]{\noalign{\vskip 6pt}\cline{#1}\noalign{\vskip 6pt}}

\newenvironment{model}
  {\itshape\list{}{\leftmargin=.5em \rightmargin=.5em
             		\parsep=0.5\baselineskip}
     \item\relax}
  {\endlist}

\input{symbols.tex}

\lehead{Ingo Nitschke and Jan Magnus Sischka and Axel Voigt}
\rohead{Hydrodynamic liquid crystal models for lipid bilayers}

\title{Hydrodynamic liquid crystal models for lipid bilayers}

\author[1,$\ast$]{Ingo Nitschke}
\author[1]{Jan Magnus Sischka}
\author[1,2,3]{Axel Voigt}
\affil[1]{Institut f{\"u}r Wissenschaftliches Rechnen, Technische Universit{\"a}t Dresden, 01062 Dresden, Germany}
\affil[2]{Center for Systems Biology Dresden (CSBD), Pfotenhauerstr. 108, 01307 Dresden, Germany}
\affil[3]{Cluster of Excellence Physics of Life (PoL), Technische Universität Dresden, 01062~Dresden, Germany}
\affil[$\ast$]{corresponding author: ingo.nitschke(at)tu-dresden(dot)de}

\begin{document}

\maketitle

\begin{abstract}
    Coarse-grained continuous descriptions for lipid bilayers are typically based on minimizing the Helfrich energy. Such models consider the fluid properties of these structures only implicitly and have been shown to nicely reproduce equilibrium properties. Model extensions that also address the dynamics of these structures are surface \mbox{(Navier--)}Stokes--Helfrich models. They explicitly account for membrane viscosity. However, these models also usually treat the lipid bilayer as a homogeneous continuum, neglecting the molecular degrees of freedom of the lipids. Here, we derive refined models which consider in addition a scalar order parameter representing the molecular alignment of the lipids along the surface normal. Starting from hydrodynamic surface liquid crystal models, we obtain a hydrodynamic surface Landau--Helfrich model for asymmetric lipid bilayers and a surface Beris--Edwards model for symmetric lipid bilayers. The fully ordered case for both models leads to the known surface \mbox{(Navier--)}Stokes--Helfrich models. Besides more detailed continuous models for lipid bilayers, we therefore also provide an alternative derivation of surface \mbox{(Navier--)}Stokes--Helfrich models. The impact on the dynamics is demonstrated by numerical simulations.  
\end{abstract}

\section{Introduction}

\begin{figure}
\centering
\begin{tikzpicture}[node distance=12pt]
	\node(SYMORD) at (0,0) {\includegraphics[width=.41\textwidth]{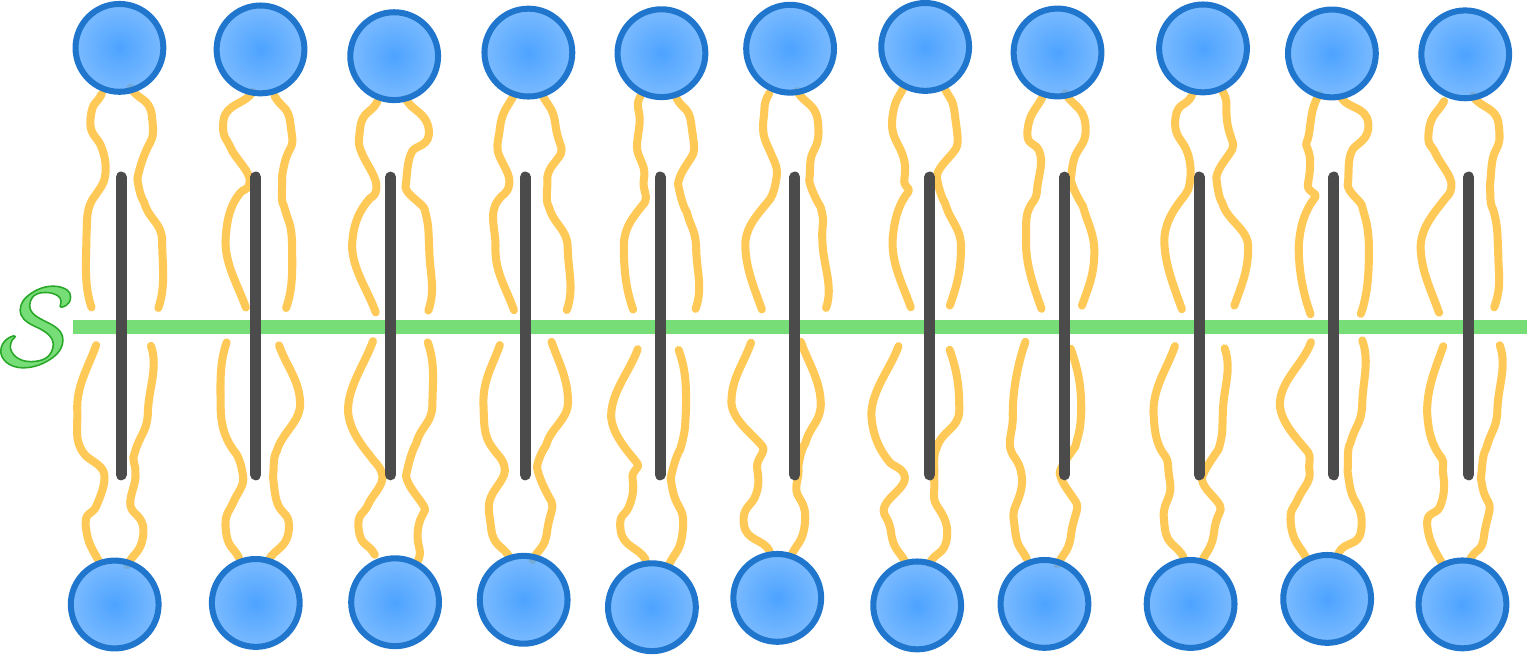}};
	\node(SYMNOTORD) [right=of SYMORD] {\includegraphics[width=.4\textwidth]{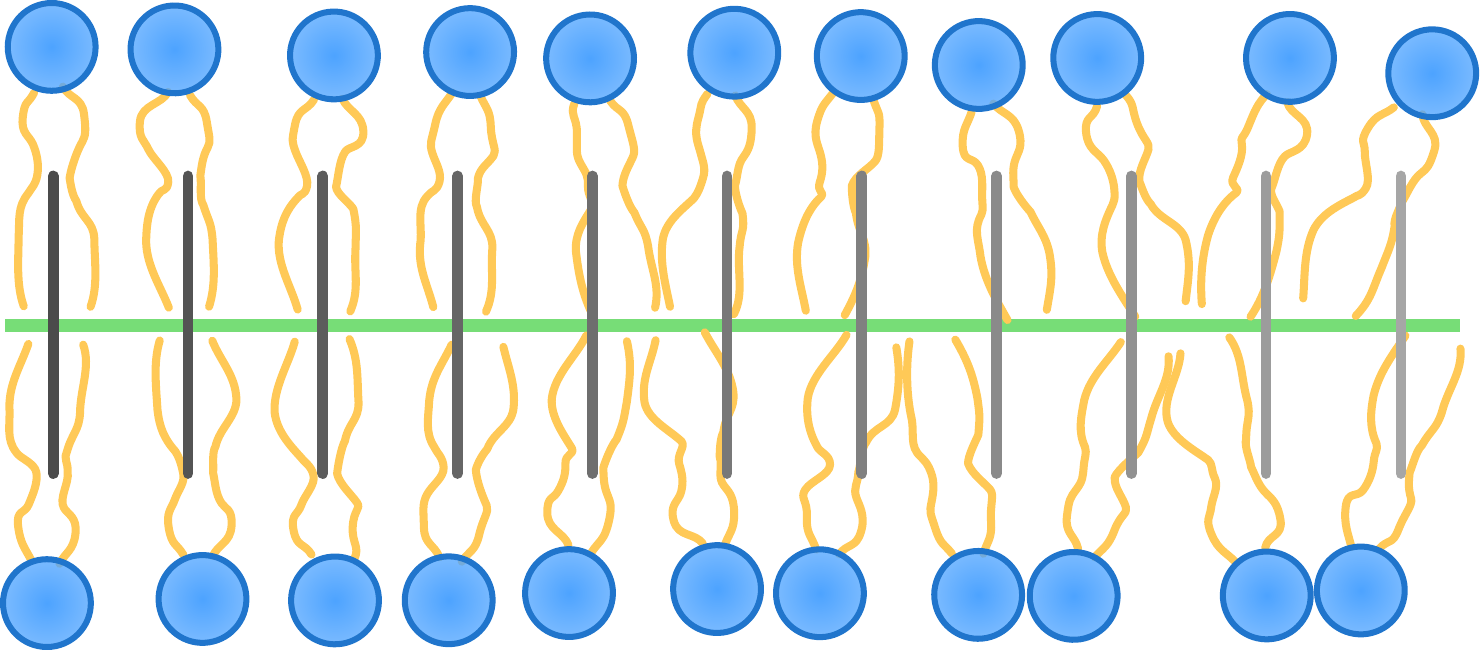}};
    \node(CB) [right=of SYMNOTORD] {\includegraphics[width=.04\textwidth]{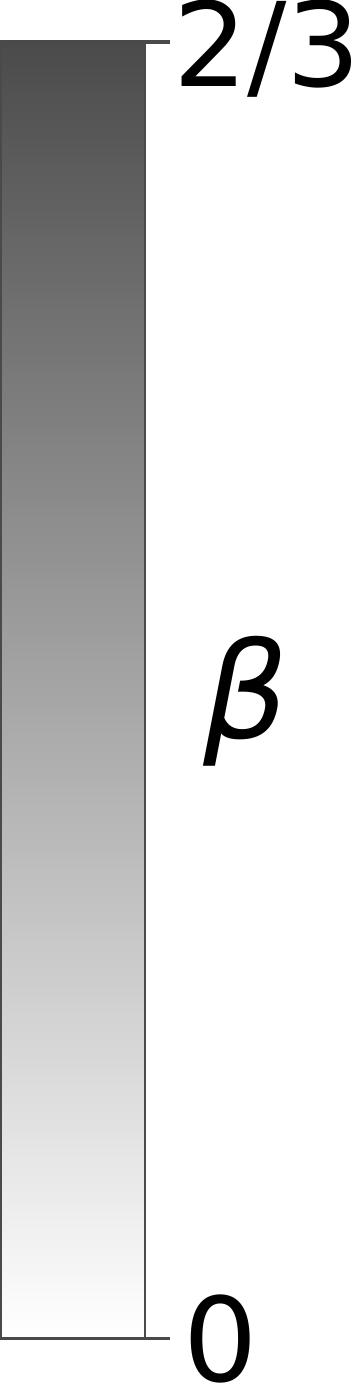}};
\end{tikzpicture}
\caption{Symmetric Lipid Bilayer.
	Left: lipid molecules are in a fully ordered state ($\beta=\frac{2}{3}$),
            \ie\ they are perfectly aligned perpendicular to the surface $ \surf $.
	Right: the degree of orientational order $\beta$ decreases from left to right, while the mean molecular alignment remains normal to the surface.
		The bilayer is represented by a surface $\surf$ (green line), 
        and the molecular orientation by a Q-tensor field $ \Qb $ fulfilling ansatz \eqref{eq:QAnsatz},
        \ie\ $ \Qb $ is depicting an apolar normal field (gray rods) with order parameter $\beta$ (grayscale). For $ \beta \neq 0 $, the lipid molecules are not in an isotropic state, the geometric minimal configuration is obtained by minimizing the Helfrich energy with zero spontaneous curvature, thus leading to a flat surface.}
\label{fig:sym_bilayers}
\end{figure}

\begin{figure}
\centering
\begin{tikzpicture}[node distance=8pt]
	\node(DC) at (0,0) {\includegraphics[width=.3\textwidth]{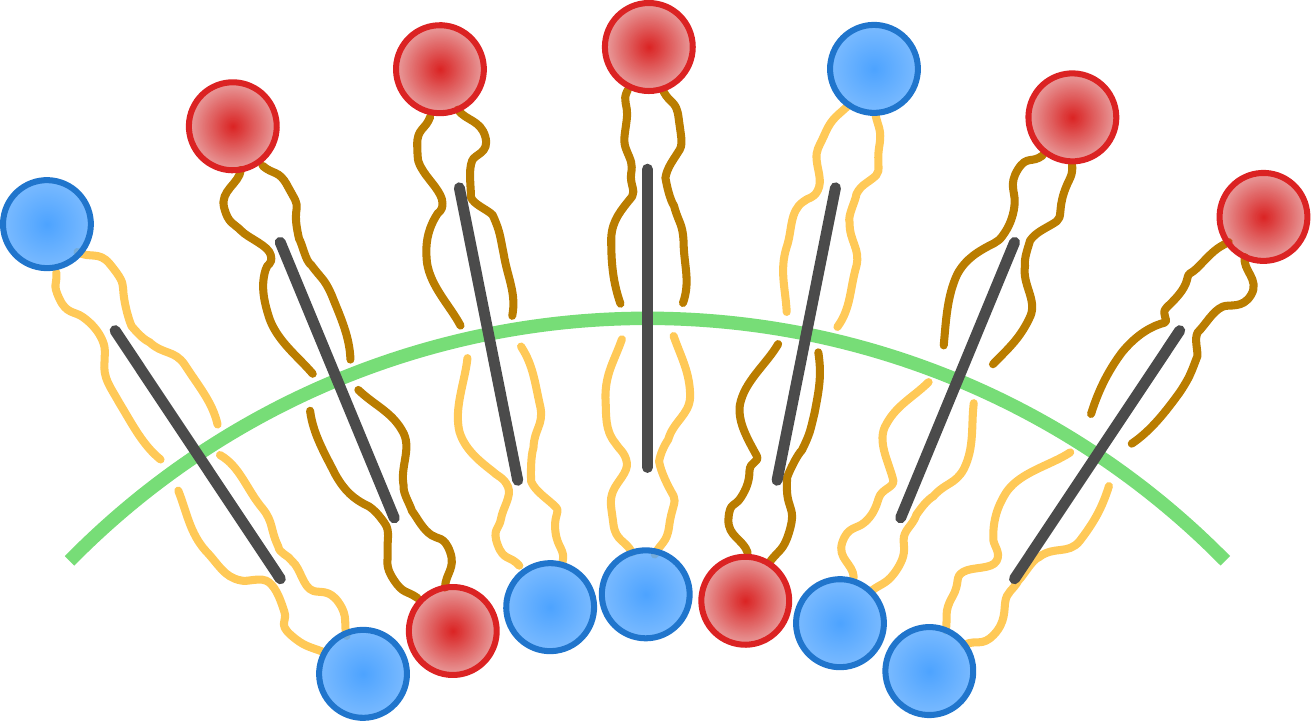}};
	\node(DA) [right=of DC] {\includegraphics[width=.3\textwidth]{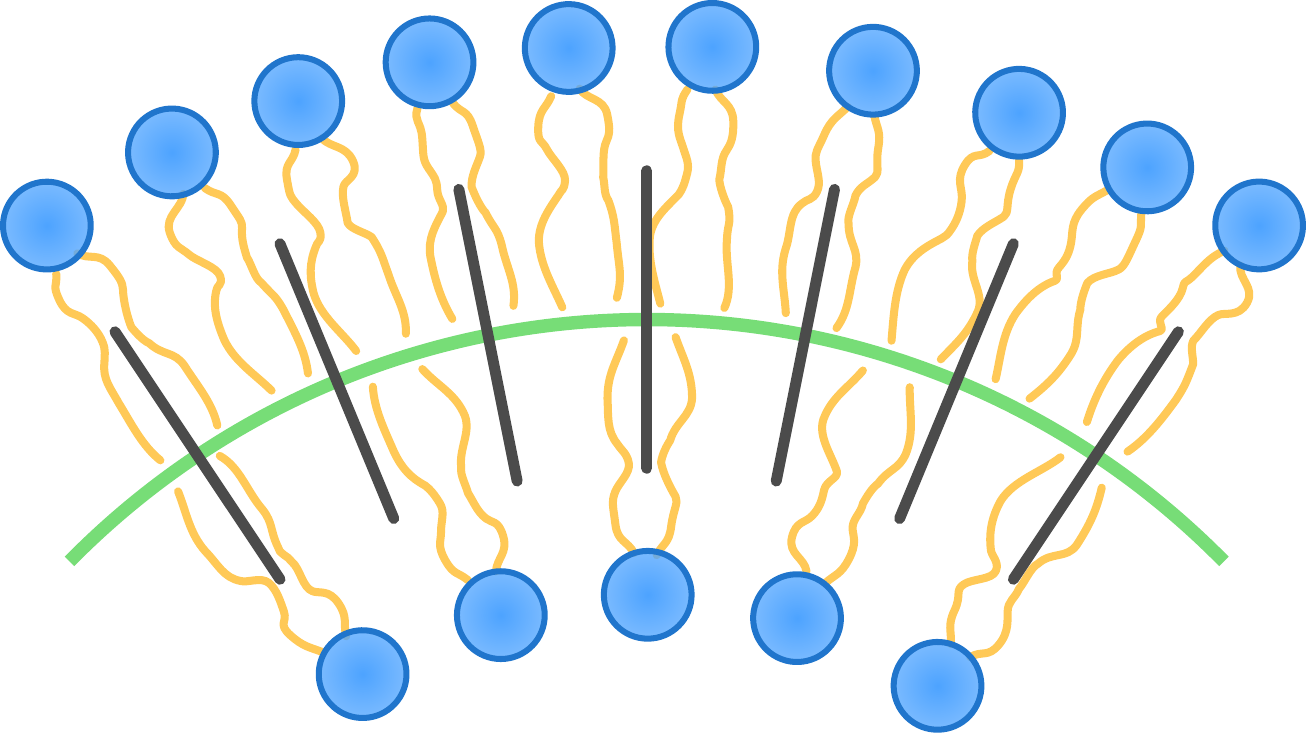}};
	\node(SC) [right=of DA] {\includegraphics[width=.3\textwidth]{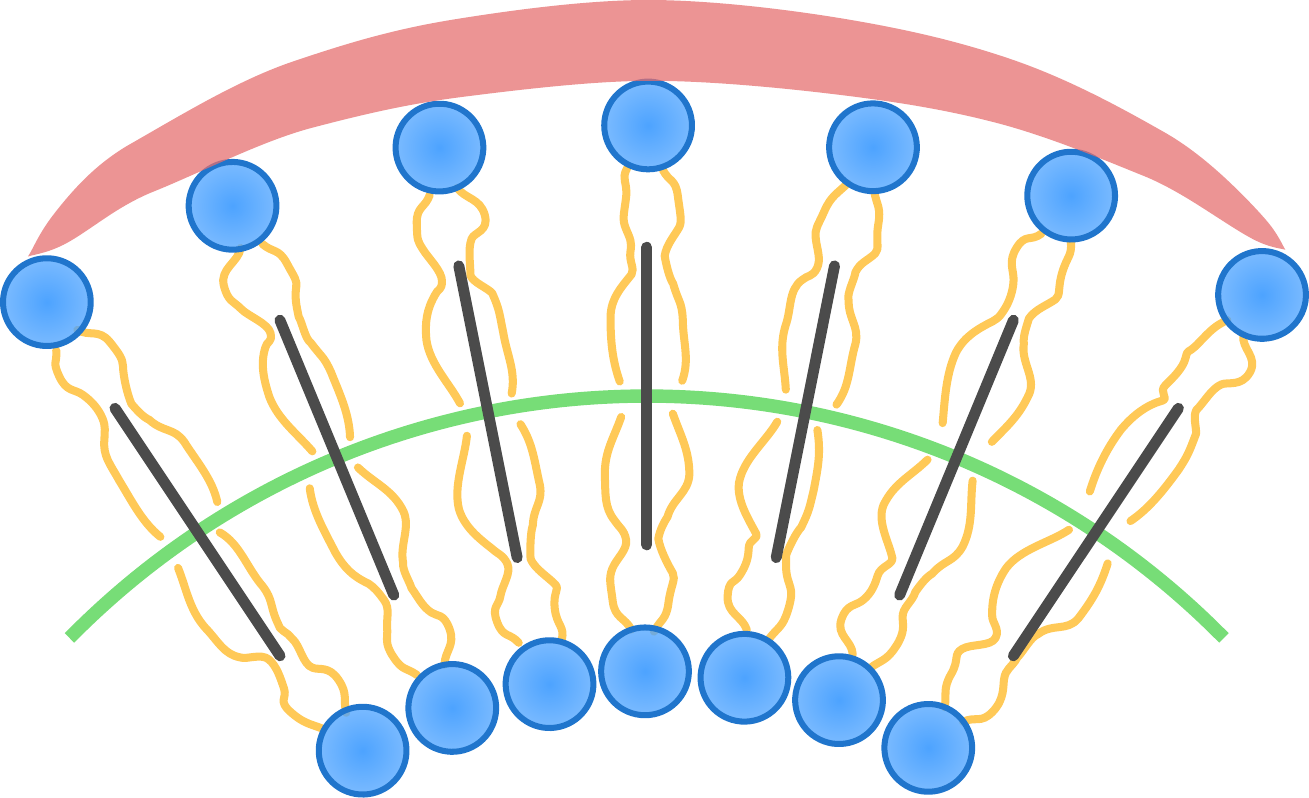}};
	\node(SC) [below=of DA, xshift=-.04\textwidth] {\includegraphics[width=.65\textwidth]{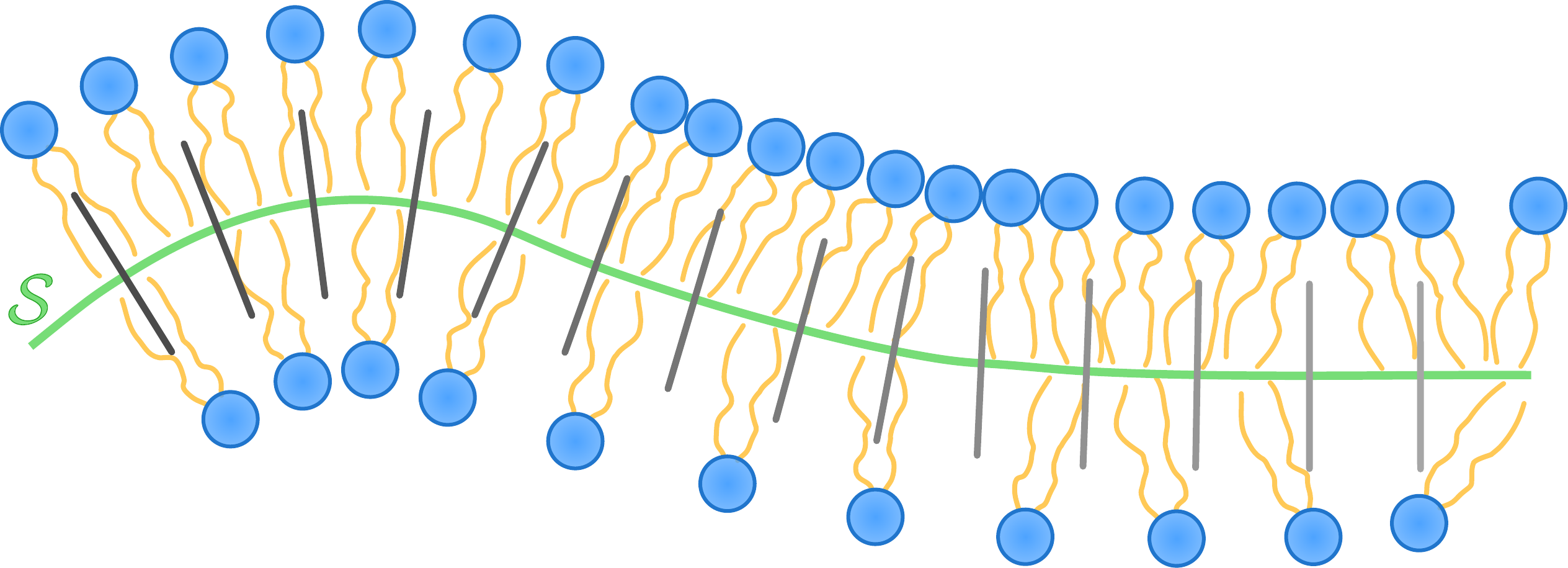}};
    \node(CB) [right=of SC, yshift=-.035\textwidth] {\includegraphics[width=.039\textwidth]{img/colorbar_order.pdf}};
\end{tikzpicture}
\caption{Asymmetric Lipid Bilayer.
		The asymmetry may arise through various mechanisms.
		We provide some examples.
		Left: differing molecular compositions.
		Center \& Bottom: differing molecular densities.
		Right: scaffold protein. The molecular orientation is represented by a Q-tensor field $ \Qb $ fulfilling ansatz \eqref{eq:QAnsatz},
        \ie\ $ \Qb $ is depicting an apolar normal field (gray rods) with order parameter $\beta$ (grayscale).
		Top: In the ordered state ($ \beta=\frac{2}{3} $), with all lipid molecules aligned normal to the surface $ \surf $ (green curve), the minimal geometric configuration is achieved when the mean curvature takes its spontaneous curvature value ($\meanc_{0}$).
		Bottom: A less ordered non-uniform state ($\beta < \frac{2}{3}$) may counteract this effect, since another spontaneous curvature ($\hat{\meanc}_0$) related to the isotropic state ($ \beta = 0 $) can be imposed additionally. The geometric minimal configuration is achieved for a curved surface with the mean curvature depending on $\beta$, $\meanc_{0}$ and $\hat{\meanc}_0$.}
\label{fig:asym_bilayers}
\end{figure}

Lipid bilayers constitute the fundamental structural components of biological membranes. They separate the cell from its exterior and organelles from the cytoplasm. In order to support life, they must balance the competing demands of fluidity versus mechanical robustness and permeability. Eukaryotic organisms solve this problem through lipid ordering. While considered in microscopic modeling approaches, continuous descriptions of lipid bilayers typically lack such information. The classical Helfrich model \citep{Helfrich_ZfNC_1973} provides the cornerstone for understanding the continuum mechanics of fluidic membranes by introducing a curvature-elastic bending energy characterized by mean and Gaussian curvature moduli and a spontaneous curvature, as well as constraints, e.g., on enclosed volume and/or global area or local inextensibility. While this model successfully captures equilibrium properties of lipid bilayers \citep{Seifert01021997}, typically used gradient flows of the energy to evolve towards the equilibrium cannot, at least not quantitatively,  describe the time-dependent, hydrodynamic evolution of biological membranes. Over the past decades, various extensions have been proposed to couple the Helfrich energy with surface fluid dynamics, typically within the framework of inextensible viscous surfaces. Such models for fluid deformable surfaces account for membrane viscosity \citep{Dimova_2006,faizi2022vesicle} and consider surface Navier--Stokes equations \citep{ReutherNitschkeEtAl_JoFM_2020, KrauseVoigt_JoCP_2023, Olshanskii_PoF_2023, SischkaNitschkeEtAl_FD_2025,Sauer_JoFM_2025}, or their Stokes limit \citep{ArroyoDeSimone_PRE_2009, Torres-SanchezMillanEtAl_JoFM_2019, ZhuSaintillanEtAl_JoFM_2025} coupled with bending properties. With this solid–fluid duality any shape change contributes to tangential flow and vice versa any tangential flow on a curved surface induces shape deformations. However, these models usually treat the biological membrane as a homogeneous continuum, neglecting the molecular degrees of freedom that can influence local curvature and flow resistance and therefore the balance between fluidity and robustness. A complementary perspective arises from liquid-crystal hydrodynamics, particularly the Beris--Edwards framework \citep{BerisEdwards_1994}, which describes the dynamics of nematic order through the evolution of a Q-tensor field coupled to hydrodynamic motion. 
Surface formulations of these models have recently been developed to study active and passive nematic layers on curved surfaces \citep{NitschkeVoigt_AiDE_2025, NitschkeVoigt_PotRSAMPaES_2025}. 
While these approaches provide a rigorous treatment of orientational order and its relaxation, they primarily apply to systems with an inherent up-down symmetry, \eg, nematic shells, or, if the average molecule directions are constrained to be perpendicular to the surface, symmetric lipid bilayers, \cf\ \cref{fig:sym_bilayers}, and thus cannot capture asymmetry characteristics of biological membranes, where the asymmetry emerges from
different molecular compositions, different molecular densities or due to scaffolding protein, \cf\ \cref{fig:asym_bilayers}. For more detailed discussions and the origin of asymmetry  in biological membranes see \citet{mcmahon2005membrane}. 

In this work, we introduce a hydrodynamic Landau--Helfrich (LH) model that bridges these two perspectives. 
Building on a polarized Landau--de Gennes energy defined on moving surfaces, we derive a self-consistent model that couples the hydrodynamics of an inextensible viscous surface to a scalar order parameter representing the molecular alignment along the surface normal. 
The model is obtained through the Lagrange--d’Alembert principle for moving surfaces \citep{NitschkeVoigt_AiDE_2025}, ensuring consistency between energy variations, viscous dissipation, and inextensibility constraints. 
The polarization of the Landau--de Gennes energy introduces curvature–order coupling terms that break up-down symmetry and thereby recover the characteristic features of asymmetric lipid bilayers, such as curvature-dependent spontaneous bending.
The resulting system consists of generalized surface Navier--Stokes equations coupled to a Landau-type relaxation equation for the order parameter. It admits several important limiting cases: for vanishing polarization, the equations lead to a special case of the surface Beris--Edwards models \citep{NitschkeVoigt_AiDE_2025}, which allows to model symmetric lipid bilayers, and for constant order, the equations reduce to the surface (Navier-)Stokes-Helfrich model for fluid deformable surfaces \citep{ArroyoDeSimone_PRE_2009, Torres-SanchezMillanEtAl_JoFM_2019, ReutherNitschkeEtAl_JoFM_2020}. In any of these derived models the bending properties of the models are not specified but result from the underlying liquid crystal interactions and in the special case of the surface \mbox{(Navier--)}Stokes--Helfrich model thus differ from previous derivations. The model formulations are expressed entirely in an observer- and coordinate-invariant tensor calculus suitable for numerical discretization and analysis within standard (ALE) surface finite-element frameworks \citep{nestler2019finite,Sauer_JoFM_2025}.
We note that the dependence on the scalar order parameter prevents us from neglecting the effect of Gaussian curvature, despite its purely intrinsic nature.
By connecting geometric surface hydrodynamics with molecular ordering through a thermodynamically consistent variational framework, the Landau--Helfrich model provides a unified continuum description of asymmetric lipid bilayers that naturally incorporates curvature-order interactions, spontaneous asymmetry, and dissipative relaxation. 
This work thus extends the classical Helfrich theory into a fully hydrodynamic regime, paving the way for systematic investigations of dynamic bilayer processes such as vesicle remodeling, protein-induced curvature generation, and the coupling of molecular ordering to membrane flow. While we mainly focus on the liquid-ordered state, phase coexistence between liquid-ordered and liquid-disordered states, as in \cite{BaumgartHessEtAl_N_2003}, can also be addressed with the proposed hydrodynamic Landau--Helfrich model.

The paper is structured as follows:
In \cref{sec:notation} we give an overview of mathematical notations relevant for this work.
Various models are listed in \cref{sec:models},
comprising the surface Beris--Edwards model for symmetric lipid bilayers (\cref{sec:BE_Model}), 
the hydrodynamic surface Landau--Helfrich model for asymmetric lipid bilayers (\cref{sec:LH_Model}),
and the surface Navier--Stokes--Helfrich model (\cref{sec:Helfrich_Model}) as a special case of the previous models for fully ordered lipid bilayers.
The derivation of these models is largely carried out in \cref{sec:derivations}.
We propose a Q-tensor ansatz (\cref{sec:QAnsatz}) for lipid molecules whose average orientation is normal to the surface.
The substitution of this ansatz into the surface-conforming Beris--Edwards model (\cref{sec:monolayer}) yields a  model for symmetric lipid bilayers.
In contrast, we also formulate a polarized surface Landau--de Gennes energy for asymmetric bilayers (\cref{sec:bilayer}), which, upon variation, leads to the hydrodynamic surface Landau--Helfrich model. 
In \cref{sec:discussion} numerical results demonstrate the differences in the dynamics if compared with surface Navier-Stokes-Helfrich models. Furthermore a summary is provided, conclusions are drawn and possible model extensions are discussed. \crefrange{app:material_parameter_constrains}{sec:NoFlatDegeneracy} provide additional information.

\section{Notation and preliminaries}\label{sec:notation}

We briefly summarize the most important notation relevant for this paper. We assume that the time-dependent moving surface $ \surf\subset\R^3 $ is sufficiently smooth and parameterizable into the 3-dimensional~Euclidean space.
Tensor fields in $ \tangentR[^n] $ are considered exclusively on the surface. 
Coordinate invariance, and the resulting freedom in the choice of frame, allow for an interpretation as a Cartesian frame for the tensor fields.
For instance, we could write $ \Wb = W^A\eb_A $ (Einstein summation) with $ A\in\{x,y,z\} $ for a vector, \resp\ 1-tensor, field $ \Wb\in\tangentR := \tangentR[^1] $.
A special vector field is the normal field $ \normal\in\tangentR $ perpendicular to the surface and normalized. 
It spans a one-dimensional subvector field space, whose orthogonal complement consists of the tangential vector fields $ \tangentS < \tangentR $.
The associated projection is given by the surface identity tensor field $ \IdS $, \ie\ it is $\tangentS = \IdS\tangentR  $ valid.
This concept readily scales to $n$-tensor fields $ \tangentR[^n] $ and tangential $n$-tensor fields $ \tangentS[^n] < \tangentR[^n] $.
The surface identity tensor field $ \IdS\in\tangentS[^2] $ is such a tangential tensor field and could be represented  by $ \IdS = \Id - \normal\otimes\normal $,
or the metric tensor of an arbitrary local tangential frame, where $ \Id\in\tangentR[^2] $ is the usual Euclidean identity tensor field, \eg\ given by $ \delta^{AB}\eb_A\otimes\eb_B $ \wrt\ a Cartesian frame.
We repeatedly make use of the corresponding orthogonal decompositions.
Force fields $ \Fb = \fb + \fnor\normal\in\tangentR $, for instance, are decomposable into a tangential force field $ \fb=\IdS\Fb\in\tangentS $ and a scalar-valued normal force field $ \fnor=\normal\Fb\in\tangentScal $,
or, stress fields
\begin{align}\label{eq:Sigmab}
     \Sigmab &= \sigmab + \normal\otimes\etab \in\tangentR\otimes\tangentS < \tangentR[^2]
\end{align}
are usually right-tangential and decomposed into a tangential stress field $ \sigmab=\IdS\Sigmab\in\tangentS[^2] $ and a vector-valued normal(-tangential) stress field $ \etab=\normal\Sigmab\in\tangentS $.
Furthermore, our framework also involves Q-tensor fields in $ \tangentQR := \{ \Qb\in\tangentR[^2] \mid \Qb = \Qb^T \text{ and } \Tr\Qb=0 \} $
and tangential Q-tensor fields in $ \tangentQS := \{ \qb\in\tangentS[^2] \mid \qb = \qb^T \text{ and } \Tr\qb=0 \}\le \tangentQR $, also known as flat-degenerated Q-tensor fields.
A complete orthogonal decomposition of $ \tangentQR $ is given in \eqref{eq:QDecomposition}.

The local inner product is denoted by angle brackets: 
$ \inner{\mathcal{V}}{\cdot,\cdot}: \mathcal{V}\times\mathcal{V} \rightarrow\tangentScal $,
where $ \mathcal{V} \le \tangentR[^n] $.
A specification of the frame is not needed as coordinate invariance and the properties of subvector spaces guarantee that the results remain consistent.
For instance, it is $ \inner{\tangentS}{\vb,\wb} =  \inner{\tangentR}{\vb,\wb} = \vb\wb = g_{ij}v^iw^j = v^i w_i = \delta_{AB}v^Aw^B$ valid for all $ \vb,\wb\in\tangentS $.
The global inner product is defined in terms of the corresponding local one: 
$ \innerH{\mathcal{V}}{\cdot,\cdot}:= \int_{\surf}\inner{\mathcal{V}}{\cdot,\cdot}\dS : \mathcal{V}\times\mathcal{V} \rightarrow\R $.
All norms are defined as those induced by the inner products: 
$ \normsq{\bullet}{\cdot} := \inner{\bullet}{\cdot,\cdot} $.

We use the componentwise derivative $ \nablaC: \tangentR[^n] \rightarrow \tangentR[^n]\otimes\tangentS < \tangentR[^{n+1}] $ as the starting point for derivatives on tensor fields, since it captures all possible first-order derivative information \wrt\ the ambient Euclidean space. 
In principle, it is the ordinary $ \R^3 $-derivative modulo the normal derivative, \ie\ $ \nablaC = (\IdS\nabla_{\R^3})\vert_{\surf} $, should one wish to consider extensions in the normal direction, although $ \nablaC $ is invariant \wrt\ any sufficiently smooth normal extension.
If we use an arbitrary tangential frame $ \{\partial_i\para\mid i=1,2\} $ we could define $ \nablaC\Rb := g^{ik}(\partial_k\Rb)\otimes\partial_i\para $ for all $ \Rb\in\tangentR[^n] $,
where $ g^{ik} $ yields the contravariant proxy of the metric tensor, also known as inverse metric tensor, \wrt\ the chosen frame.
Equivalently, it holds $ [\nablaC\Rb]^{A_1 \ldots A_n B} \eb_B = \nabla R^{A_1 \ldots A_n}$ for $ \Rb= R^{A_1 \ldots A_n} \eb_{A_1}\otimes\ldots\otimes \eb_{A_n}$ in a Cartesian frame and covariant derivative $ \nabla $ on scalar fields.
The shape operator $ \shop\in\tangentS[^2] $, also known as the second fundamental form, follows directly from this via $ \shop = -\nablaC\normal $.
Other curvature related quantities derived from this are the mean curvature $ \meanc=\Tr\shop = \IdS\dbdot\shop $ 
and the Gaussian curvature $ \gaussc = \frac{1}{2}\left( \meanc^2 - \normsq{\tangentS[^2]}{\shop} \right) $, the two scalar-valued invariants of the shape operator.
With respect to a local tangential frame, it holds also $ \gaussc = \det\{\shopC^i_j\} $, or $ \gaussc = \frac{1}{4}\Eb\dbdot\boldsymbol{\mathcal{R}}\dbdot\Eb $, where $ \Eb\in\tangentS[^2] $ is the Levi-Civita and $ \boldsymbol{\mathcal{R}}\in\tangentS[^4] $ is the Riemann curvature tensor field.
On scalar fields the componentwise and covariant derivative are the same, \ie\ $ \nablaC=\nabla:\tangentScal\rightarrow\tangentS $.
This is not generally valid, not even for tangential tensor fields.
For a vector field $ \Vb=\vb+\vnor\normal\in\tangentR $, \eg\, holds
$ \nablaC\Vb = \nabla\vb - \vnor\shop + \normal\otimes\left( \nabla\vnor + \shop\vb\right) $.
As the divergence operator, we use the component-wise trace divergence $ \DivC := \Tr\circ\nablaC $, where the trace applies on the two rear column dimensions, including the derivative acting on the last index.
It should be noted that the trace divergence equals the $ \hil $-adjoint $ -\nablaC^{*} $ only for right-tangential tensor fields,
since the correct relation is $ \DivC\Rb = - \left( \nablaC^{*}\Rb + \meanc\Rb\normal \right) $ for all $ \Rb\in\tangentR[^n] $.
Fortunately, this is true for fluid stress fields \eqref{eq:Sigmab} in this paper,
\eg\ it holds $ \innerH{\tangentR}{\DivC\Sigmab, \Wb} = -\innerH{\tangentR[^2]}{\Sigmab, \nablaC\Wb} $ in that case for all $ \Wb\in\tangentR $.
However, at least for pressure gradients, we need the $ \hil $-adjoint of the trace divergence.
This gives rise to the adjoint gradient $ \GradC:=-\DivC^{*} $, which yields the relation $ \GradC\Rb= \nablaC\Rb + \meanc\Rb\otimes\normal $ for all $ \Rb\in\tangentR[^n] $ as a consequence.
The key representations in the covariant differential calculus relevant to this paper are 
$ \DivC\Sigmab = \div\sigmab - \shop\etab + \left( \div\etab + \shop\dbdot\sigmab \right)\normal $
for right-tangential fields \eqref{eq:Sigmab},
$ \DivC\Vb = \div\vb - \meanc\vnor $
for vector fields $ \Vb=\vb+\vnor\normal\in\tangentR $,
and $ \GradC p = \DivC(p \IdS) = \nabla p + p \meanc \normal $
for scalar fields $ p\in\tangentScal $.

Even if our focus is not on the numerical solution of the proposed models, we consider $ \R^3 $-representations with componentwise differential operators as they are the most convenient for standard numerical implementations, e.g., by the surface finite element method. First, the operators act directly on the Cartesian proxy of the material velocity (\eg\ $ \DivC\Vb = \left( \delta^B_A-\normalC_A \normalC^B \right)\partial_B V^A $ for $ A,B\in\{x,y,z\} $). Second, differential operators straightforwardly admit a weak formulation due to the $ \hil $-adjoint relations 
\begin{align*}
    \innerH{\tangentR}{\DivC\Sigmab, \Wb}
        &= -\innerH{\tangentR[^2]}{ \Sigmab, \nablaC\Wb} \formComma\\
    \innerH{\tangentScal}{\DivC\Wb, \psi}
        &= - \innerH{\tangentR}{\Wb, \GradC\psi}
\end{align*}
for all $ \Sigmab \in \tangentR\otimes\tangentS $, $ \Wb\in\tangentR $, and $ \psi\in\tangentScal $. For a more comprehensive overview on tensor calculus on moving surfaces, see the introduction chapter in \citet{Nitschke_2025}, or the detailed mathematical introductions with respect to time derivatives in \citet{NitschkeVoigt_JoGaP_2023}, with respect to energy variations in \citet{NitschkeSadikVoigt_IJoAM_2023}, and  with respect to the Lagrange--d’Alembert principle in \citet{NitschkeVoigt_AiDE_2025}.

\section{Models}\label{sec:models}

\subsection{Surface Beris--Edwards model for symmetric lipid bilayers}\label{sec:BE_Model}

A hydrodynamic liquid crystal model for symmetric lipid bilayers can be obtained from the surface Beris--Edwards (BE) framework \citep{NitschkeVoigt_AiDE_2025}.
We carry out this derivation in \cref{sec:monolayer} starting from the surface conforming Beris--Edwards model \citep{NitschkeVoigt_AiDE_2025} (Sec. 3.3), 
conveniently provided in \cref{sec:full_surface_conforming_models}.
This leads to: \\

\begin{model}
Find the material velocity field $ \Vb\in\tangentR $, scalar field $ \beta\in\tangentScal $
and generalized pressure field $ \tp\in\tangentScal $ \soth
\begin{subequations}\label{eq:BE_R3}
\begin{gather}
    \rho\left( \partial_t\Vb + (\nablaC\Vb)\left( \Vb - \Vb_{\!\ofrak} \right) \right) \label{eq:BE_fluid}
        = \GradC\tp + \DivC\tSigmab\\
   \begin{align}
      \text{with }\quad\tSigmab
      	&= \coeffIF\left( 1 + \frac{\xi}{2}\beta \right)^2 \left( \IdS\nablaC\Vb + (\IdS\nablaC\Vb)^T \right)
      	                        - \frac{3L}{2}\left( \nablaC\beta \otimes \nablaC\beta + 3\beta^2\meanc\shop \right) \notag\\
      	&\quad+ \frac{9}{2}\normal\otimes\left( \delta_{\mfrak}^{\Phi} M \beta^2 \normal\nablaC\Vb
      	                                       - L\beta\left( \beta\nablaC\meanc + 2\shop\nablaC\beta \right)  \right) \notag
    \end{align}\\
    \begin{align}
    	\left( M + \frac{\coeffIF\xi^2}{2} \right) \left( \partial_t\beta + (\nablaC\beta)\left( \Vb - \Vb_{\!\ofrak} \right) \right) \label{eq:BE_mol} \hspace{-7em}\\
        	&= L\left(\Delta_{\Comp}\beta - 3\beta\left( \meanc^2 - 2\gaussc \right)\right) 
            	-\left(2a + b\beta + 3c\beta^2\right)\beta \notag
    \end{align}\\
    \DivC\Vb \label{eq:BE_konti}
        = 0
\end{gather}
\end{subequations}
holds for $ \dot{\rho} = 0 $ and given initial conditions for $ \Vb $, $ \beta $, and mass density $ \rho\in\tangentScal $. \\
\end{model}

$ \Vb_{\!\ofrak}\in\tangentR $ is the so-called observer velocity, which is arbitrary, not necessarily divergence-free, and could be used as mesh velocity in a discretized problem for instance.
We use $ \beta\in\tangentScal $ as a proxy for the scalar order parameter $ S\in\tangentScal $. 
In fact, $ \beta $ is the eigenvalue of the corresponding Q-tensor field in normal direction. 
Ultimately, $ \beta $ statistically quantifies the degree to which the lipid molecules are aligned along the normal direction: 
$ \beta = \frac{2}{3} $ for the fully ordered state and $ \beta=0 $ for the isotropic, \ie\ disordered, state.
In this process, the average molecular orientation is constant along the normal direction.
The fluid stress tensor $\tSigmab\in\tangentR\otimes\tangentS$ is given modulo $ \IdS $; 
the resulting pressure gradient is, respectively, incorporated into the generalized pressure $ \tp $.
Material parameters are the isotropic viscosity $\coeffIF $, anisotropy coefficient $\xi$, elastic parameter $L$, immobility coefficient $M$, and thermotropic coefficients $a$, $b$, and $c$, \cf\ \cref{tab:material_parameter}.
The Kronecker--delta $\delta_{\mfrak}^{\Phi}$ acts as a switch that selects between the material model ($\Phi=\mfrak$: $ \delta_{\mfrak}^{\Phi} = 1$) and the Jaumann model ($\Phi=\jau$: $ \delta_{\mfrak}^{\Phi} = 0$).
This distinction is less relevant here than in the more general Beris--Edwards models \citep{NitschkeVoigt_AiDE_2025}.
However, the resulting material immobility force accounts for changes of orientations of the lipids due to the motion of the surface with respect to the surrounding 3-dimensional space. 
This may become relevant if one intends to include the mass, and hence local torque, of the lipid molecules. 
The Jaumann model, on the other hand, is invariant under rigid-body rotations; see the discussion in \citet{NitschkeVoigt_AiDE_2025}.
In any case, for slow and small lipid surfaces, the choice of model will hardly lead to any qualitative difference in the solution. The lack of a spontaneous curvature term reflects the presence of the up–down symmetry of the surface,
something that the general surface Beris--Edwards model in \citet{NitschkeVoigt_AiDE_2025} already exhibits by construction. As a consequence the model is appropriate for symmetric lipid bilayers. 

In \citet{NitschkeVoigt_PotRSAMPaES_2025} additional active mechanisms are considered in the derivation of the general surface Beris--Edwards model. These terms do not contribute in the present setting due to inextensibility and the orthogonal alignment of the lipid molecules with respect to the surface. As a consequence, even with these terms, the solutions exhibit a purely dissipative energy evolution.

The governing equations \eqref{eq:BE_R3} can be transformed into a tangential calculus in the usual manner. 
The tangential and normal fluid forces on the right-handed side of \eqref{eq:BE_fluid} yield
\begin{align*}
    \IdS\left(\GradC\tp + \DivC\Sigmab\right) \hspace{-7em}&\\
    	&= \nabla\hat{p}
    	   +\coeffIF \left(1+\frac{\xi}{2}\beta\right)^2  
                    \left(\Delta\vb + \gaussc\vb - 2\left( \shop - \frac{\meanc}{2}\IdS \right)\nabla\vnor 
    	         	         - \vnor\nabla\meanc\right)\\
    	&\quad+\coeffIF \xi \left(1+\frac{\xi}{2}\beta\right)\left( \nabla\vb + \nabla^T\vb - 2\vnor\shop \right)\nabla\beta 
    	    -\frac{3L}{2} \left( \Delta\beta  - 3\beta \left( \meanc^2 - 2\gaussc \right) \right) \nabla\beta\\
    	&\quad +\frac{9 M}{2} \delta_{\mfrak}^{\Phi} \beta^2\left( \gaussc\vb - \shop\left( \nabla\vnor + \meanc\vb \right) \right) \formComma\\
    \normal\left(\GradC\tp + \DivC\Sigmab\right) \hspace{-7em}&\\
    	&= \meanc\hat{p}
    	    +2\coeffIF \left(1+\frac{\xi}{2}\beta\right)^2 \left( \shop\dbdot\nabla\vb - \vnor\left( \meanc^2 - \gaussc \right) \right) 
    		-\frac{3L}{4}\left( 14\nabla\beta\shop\nabla\beta - \meanc\normsq{}{\nabla\beta}\right)\\
    	 &\quad -\frac{9L}{4}\beta\left( 4\shop\dbdot\nabla^2\beta + 8\nabla\meanc\nabla\beta 
    	                                 +\beta\left( 2\Delta\meanc + \meanc(\meanc^2 - 4\gaussc) \right)\right)\\
    	 &\quad +\frac{9 M}{2} \delta_{\mfrak}^{\Phi} \beta \left( \left( \Delta\vnor + \shop\dbdot\nabla\vb + \nabla_{\vb}\meanc \right)\beta 
    	     									+ 2\left( \nabla\vnor + \vb\shop \right)\nabla\beta\right) \formComma
\end{align*}
where $\Vb = \vb + \vnor\normal$, see \cref{sec:monolayer}.
For purely aesthetic reasons, we have separated the elastic pressure contribution from the pressure $\tp$ again, 
i.e., $\tp = \hat{p} + \frac{3L}{4}\left( \normsq{}{\nabla\beta} + 3\meanc^2\beta^2 \right)$.
The tangential and normal material accelerations forces on the left-handed side of \eqref{eq:BE_fluid} are  
\begin{align}\label{eq:accelaration_split}
\begin{aligned}
    \rho\ab 
        &= \rho\IdS\left( \partial_t\Vb + (\nablaC\Vb)\left( \Vb - \Vb_{\!\ofrak} \right) \right)\\ 
    	&= \rho\left((\partial_t v^i)\partial_i\para_{\!\ofrak} + \nabla_{\vb-\vb_{\!\ofrak}}\vb + \nabla_{\vb}\vb_{\!\ofrak} - \vnor\left( \nabla\vnor + 2\shop\vb \right)\right) \formComma\\
   	\rho\anor 
        &= \rho\normal\left( \partial_t\Vb + (\nablaC\Vb)\left( \Vb - \Vb_{\!\ofrak} \right) \right)\\
   		&= \rho\left(\partial_t\vnor + \nabla_{2\vb-\vb_{\!\ofrak}}\vnor + \shop(\vb,\vb)\right) \formComma
\end{aligned}
\end{align}
where $\Vb_{\!\ofrak} = \vb_{\!\ofrak} + \vnor\normal $, and $\{\partial_i\para_{\!\ofrak}\}$ is the observer frame, see \cref{tab:forces_conforming}. The inextensibility constraint \eqref{eq:full_R3_konti} becomes: $ \div\vb = \vnor\meanc $. 
The fully ordered case ($\beta=\frac{2}{3}$) leads to a surface Navier-Stokes-Helfrich model and is discussed in \cref{sec:Helfrich_Model}.

\subsection{Hydrodynamic surface Landau--Helfrich model for asymmetric lipid bilayers}\label{sec:LH_Model}

The hydrodynamic surface Landau--Helfrich (LH) model for asymmetric lipid bilayers is derived in \cref{sec:variations} following the Lagrange-D'Alembert principle, and energetic considerations in \cref{sec:bilayer} and \citet{NitschkeVoigt_AiDE_2025}.
The model can be written as: \\

\begin{model}
Find the material velocity field $ \Vb\in\tangentR $, scalar field $ \beta\in\tangentScal $
and generalized pressure field $ \tp\in\tangentScal $ \soth
\begin{subequations}\label{eq:full_R3}
\begin{gather}
    \rho\Dmat\Vb \label{eq:full_R3_fluid}
        = \GradC\tp
           +\DivC\left( \tsigmabNV + \tsigmab_1 + \tsigmab_0
                        + \normal\otimes\left( \delta_{\mfrak}^{\Phi}\etab_{\IM}^{\mfrak} + \etab_0 + \etab_{\bar{\kappa}}\right)\right)\\
    \left( M + \frac{\coeffIF\xi^2}{2} \right) \dot{\beta} \label{eq:full_R3_mol}
        = \omega_1
            + \omega_0 
            + \omega_{\bar{\kappa}} 
            + \omega_{\text{dw}}\\
    \DivC\Vb \label{eq:full_R3_konti}
        = 0
\end{gather}
\end{subequations}
holds for $ \dot{\rho} = 0 $ and given initial conditions for $ \Vb $, $ \beta $, and mass density $ \rho\in\tangentScal $. \\
\end{model}

Quantities for the fluid equation \eqref{eq:full_R3_fluid} are listed in \cref{tab:quantities_full_R3_fluid},
and for the molecular equation \eqref{eq:full_R3_mol} in \cref{tab:quantities_full_R3_mol}.
Material parameters can be found in \cref{tab:material_parameter}.
Again the stress tensors are given modulo $ \IdS $; 
the resulting pressure gradients are, respectively, incorporated into the generalized pressure $ \tp $. Analogous to the surface Beris--Edwards model \eqref{eq:BE_R3}, activity, if additionally considered as in \citet{NitschkeVoigt_PotRSAMPaES_2025}, has no effect, 
\ie\ also the solutions of the hydrodynamic surface Landau--Helfrich model exhibit a purely dissipative energy evolution as we can see in \cref{sec:energy_rate}.
Many of the remarks made in \cref{sec:BE_Model} remain valid in this context as well and are therefore not reiterated here,
since \eqref{eq:BE_R3} and \eqref{eq:full_R3} coincide for
\begin{align}\label{eq:consistency}
\begin{aligned}
     a &= \frac{2\hat{a}}{27} \formComma
    &b &= -\frac{2\hat{a}}{3} - 27\varpi \formComma
    &c &= \frac{2\hat{a}}{9} + \frac{27\varpi}{2} \formComma\\
     \meanc_{0} = \hat{\meanc}_0 &= 0 \formComma
    &\kappa=\bar{\kappa} &= 2L \formPeriod
\end{aligned}
\end{align}

\begin{table}
    \centering
    \begin{tabular}{lcr}
            Identifier
                & Expression
                    & \\
        \hline
            $ \Dmat\Vb $
                & $ \partial_t\Vb + (\nablaC\Vb)\left( \Vb - \Vb_{\!\ofrak} \right) $
                    & material acceleration \citep{NitschkeVoigt_JoGaP_2023}\\
        \hline
            $ \tsigmabNV $
                & $ \coeffIF\left( 1 + \frac{\xi}{2}\beta \right)^2 \left( \IdS\nablaC\Vb + (\IdS\nablaC\Vb)^T \right) $
                    & deviatoric nematic viscous stress \eqref{eq:NV_force_stress}\\
        \hline
            $ \tsigmab_1 $
                & $ -\frac{3L}{2} \nablaC\beta\otimes\nablaC\beta $
                    & part of the LH stress \eqref{eq:LH1_force_stress}\\
        \hline
            $  \tsigmab_0 $
                & $ - \frac{9\kappa}{4}\beta^2\left( \meanc - \hat{\meanc}_{0} - \frac{3}{2} \left( \meanc_{0} - \hat{\meanc}_{0} \right)\beta \right) \shop $
                    & part of the LH stress \eqref{eq:LH0_force_stress}, \eqref{eq:LH0_force_stress_expanded}\\
        \hline
            $ \etab_{\IM}^{\mfrak} $
                & $ \frac{9 M}{2} \beta^2 \normal\nablaC\Vb $
                    & material immobility normal stress \eqref{eq:IM_force_stress}\\
        \hline
            $ \etab_0 $
                & $ - \frac{9\kappa}{4}\nablaC\left(\beta^2\left( \meanc - \hat{\meanc}_{0} - \frac{3}{2} ( \meanc_{0} - \hat{\meanc}_{0} )\beta \right) \right) $
                    & part of the LH normal stress \eqref{eq:LH0_force_stress}, \eqref{eq:LH0_force_stress_expanded}\\
        \hline
            $ \etab_{\bar{\kappa}} $
                & $ \frac{9\bar{\kappa}}{2} \beta \left( \meanc\IdS - \shop \right) \nablaC\beta $
                    & part of the LH normal stress \eqref{eq:LHbarkappa_stress}
    \end{tabular}
    \caption{Quantities appearing in the fluid equation \eqref{eq:full_R3_fluid}.
                $ \Vb_{\!\ofrak}\in\tangentR $ is the so-called observer velocity, which is arbitrary, not necessarily divergence-free, and could be used as mesh velocity in a discretized problem for instance.
                Stress tensors $ \tsigmab $ are written modulo $ \IdS $, \resp\ resulting pressure gradient, in comparison to the listed references, 
                since the model \eqref{eq:full_R3} is inextensible.
                The use of $ \nablaC $ on $ \tangentScal $ instead of $ \nabla $ is merely cosmetic; both are equivalent on scalar fields.
                Note that $ \etab_{\IM}^{\mfrak} $ applies only within the material model.}
    \label{tab:quantities_full_R3_fluid}
\end{table}

\begin{table}
    \centering
    \begin{tabular}{lcr}
            Identifier
                & Expression
                    & \\
        \hline
            $ \dot{\beta} $
                & $ \partial_t\beta + (\nablaC\beta)\left( \Vb - \Vb_{\!\ofrak} \right) $
                    & $ \beta $ rate \citep{NitschkeVoigt_JoGaP_2023}\\
        \hline
            $ \omega_1 $
                & $  L \Delta_{\Comp}\beta $
                    & part of the mol.\@ LH force \eqref{eq:LH1_force_stress}\\
        \hline
            $ \omega_0 $
                & $ -\frac{3}{2}\kappa \beta\left( \meanc - \hat{\meanc}_{0} - \frac{3}{2} ( \meanc_{0} - \hat{\meanc}_{0} )\beta \right)
                                              \left( \meanc - \hat{\meanc}_{0} - 3 ( \meanc_{0} - \hat{\meanc}_{0} )\beta \right) $
                    &  part of the mol.\@ LH force \eqref{eq:LH0_force_stress_expanded}\\
        \hline
            $ \omega_{\bar{\kappa}} $
                & $ 3\bar{\kappa} \gaussc \beta $
                    & part of the mol.\@ LH force \eqref{eq:LHbarkappa_force}\\
        \hline
            $ \omega_{\text{dw}} $
                & $ -\frac{1}{6}\beta\left( \frac{2}{3} - \beta \right)\left( 4 \hat{a}\left( \frac{1}{3}-\beta \right)-243\varpi\beta \right)  $
                    & part of the mol.\@ LH force \eqref{eq:LHdw_stress_force}
    \end{tabular}
    \caption{Quantities appearing in the molecular equation \eqref{eq:full_R3_mol}.
                $ \Vb_{\!\ofrak}\in\tangentR $ is the so-called observer velocity, which is arbitrary, not necessarily divergence-free, and could be used as mesh velocity in a discretized problem for instance.
                The use of component-wise operators $ \nablaC, \Delta_{\Comp} $ on $ \tangentScal $ instead of covariant $ \nabla, \Delta $ is merely cosmetic; both are equivalent on scalar fields.}
    \label{tab:quantities_full_R3_mol}
\end{table}

\begin{table}
    \centering
    \begin{tabular}{lllr}
            Parameter
                & Domain
                    & Name
                        & Origin\\
        \hline
            $ \coeffIF $
                & $ \ge 0 $
                    & isotropic viscosity
                        & aniso.\@ viscous flux potential \eqref{eq:visc_flux_pot}\\
        \hline
            $ \xi $
                & $ \in \R $
                    & anisotropy coefficient
                        & aniso.\@ viscous flux potential \eqref{eq:visc_flux_pot}\\
        \hline
            $  M $
                & $ \ge 0 $
                    & immobility coefficient
                        & immobility flux potential \eqref{eq:immobility_flux_pot}\\
        \hline
            $ L $
                & $ \ge 0 $
                    & elastic parameter
                        & LH potential energy \eqref{eq:LH}\\
        \hline
            $\meanc_{0}$
                & $ \in\R $
                    & spontaneous curvature at ordered state
                        &  LH potential energy \eqref{eq:LH}\\
        \hline
            $ \hat{\meanc}_{0} $
                & $ \in\R $
                    & spontaneous curvature at isotropic state 
                        & LH potential energy \eqref{eq:LH}\\
        \hline
            $ \kappa $
                & $ \ge 0 $
                    & mean curvature-elastic moduli
                        & LH potential energy \eqref{eq:LH}\\
        \hline
            $ \bar{\kappa} $
                & $\in[0,\kappa]$
                    & Gaussian curvature-elastic moduli
                        & LH potential energy \eqref{eq:LH}\\
        \hline
            $ \varpi $
                & see \eqref{eq:DW_parameter_regime}
                    & double-well parameter
                        & LH potential energy \eqref{eq:LH}\\
        \hline
            $ \hat{a} $
                & see \eqref{eq:DW_parameter_regime}
                    & double-well parameter
                        & LH potential energy \eqref{eq:LH}
    \end{tabular}
    \caption{Material parameters for the LH model \eqref{eq:full_R3}.
        Given domains are only necessary, but not sufficient, for solvability or physical plausibility.
        For instance, \citet{NitschkeVoigt_AiDE_2025} suggest $ -3 < 2\xi < 3 $ to maintain a positive definite lipid metric.}
    \label{tab:material_parameter}
\end{table}

Again, we also formulate the model within a tangential calculus, including the use of covariant derivatives. For this purpose we split the fluid equation \eqref{eq:full_R3_fluid} into its projective tangential and normal part:
\begin{align*}
    \rho\ab
        &= \nabla\bp + \fb_{\NV} + \delta_\mfrak^\Phi \fb_{\IM}^{\mfrak} + \fb_1 + \fb_0 + \fb_{\bar{\kappa}} \formComma\\
    \rho\anor
        &= \bp\meanc + \fnor[\NV] + \delta_\mfrak^\Phi f_{\IM}^{\mfrak,\bot} +\fnor[1] + \fnor[0] + \fnor[\bar{\kappa}] \formComma
\end{align*}
where the tangential and normal fluid forces 
$ \fb_{\NV} $, $ \fnor[\NV] $ \eqref{eq:NV_fluid_force}, 
$ \fb_{\IM}^{\mfrak} $, $ f_{\IM}^{\mfrak,\bot} $ \eqref{eq:IM_fluid_force},
$ \fb_1 $, $ \fnor[1] $ \eqref{eq:LH1_force_stress},
$ \fb_0 $, $ \fnor[0] $ \eqref{eq:LH0_force_stress}, \eqref{eq:LH0_force_stress_expanded},
and $ \fb_{\bar{\kappa}} $, $ \fnor[\bar{\kappa}] $ \eqref{eq:LHbarkappa_force}
are given in \cref{sec:variations}.
It should be noted that
\begin{align*}
    \fb_1 + \fb_0 + \fb_{\bar{\kappa}}
        &= -\frac{3}{2}\left( \omega_1 + \omega_0 + \omega_{\bar{\kappa}} \right)\nabla\beta
\end{align*}
holds, \ie\ the tangential surface Landau--Helfrich fluid force vanishes for constant $ \beta $,
as expected, since in this case the force is purely geometric 
and the associated normal force $ \fnor[1] + \fnor[0] + \fnor[\bar{\kappa}] $ reduces to a curvature-driven Helfrich force.
Explicit representation of the tangential $ \ab = \IdS\Dmat\Vb $ and normal acceleration $ \anor = \normal\Dmat\Vb $ 
depending on an arbitrary observer frame can be found at top of \cref{tab:forces_conforming}, \resp\ \eqref{eq:accelaration_split}.
Even though this has no influence on the solution, the generalized pressure $ \bp $ is different here than in \eqref{eq:full_R3_fluid}.
We have included in $ \bp $ only the forces arising from the double-well potential \eqref{eq:LHdw_stress_force} and the activity \eqref{eq:AC_stress_force}, 
both of which reduce to pure pressure gradients.
The molecular equation \eqref{eq:full_R3_mol} can essentially be used unchanged. 
In \cref{tab:quantities_full_R3_mol} the subscript $ \Comp $ may be omitted, 
as for scalar fields the respective differential operators coincide with the covariant ones.
Within the scalar rate $ \dot{\beta} $ the relation $ \Vb - \Vb_{\!\ofrak} = \vb - \vb_{\!\ofrak} $ already holds anyway, 
since all observers for the same moving surface have the same normal velocity. 
The inextensibility constraint \eqref{eq:full_R3_konti} becomes: $ \div\vb = \vnor\meanc $. The fully ordered case ($\beta=\frac{2}{3}$) leads to a surface Navier-Stokes-Helfrich model and is discussed in \cref{sec:Helfrich_Model}.

\subsection{Surface Navier-Stokes-Helfrich models}\label{sec:Helfrich_Model}

In the vicinity of the ordered equilibrium, we may assume $ \beta=\frac{2}{3} $. 
Incorporating this assumption into model \eqref{eq:full_R3}, \eg\ by means of a Lagrange multiplier formulation, 
eliminates the molecular equation \eqref{eq:full_R3_mol} and yields the surface Navier-Stokes-Helfrich model: \\

\begin{model}
Find the material velocity field $ \Vb\in\tangentR $
and generalized pressure field $ \tp\in\tangentScal $ \soth
\begin{subequations}\label{eq:H_R3}
\begin{gather}
\begin{align}
    \rho\Dmat\Vb \label{eq:H_fluid}
        &= \GradC\tp
           +\DivC\Big( \tilde{\coeffIF}\left( \IdS\nablaC\Vb + (\IdS\nablaC\Vb)^T \right) 
                        - \kappa\left( \meanc - \meanc_0 \right)\shop\\ 
        &\hspace{8em}+ \normal\otimes\left( 2M\delta_{\mfrak}^{\Phi} \normal\nablaC\Vb
                                                - \kappa\nablaC\meanc  \right)\Big)\notag
\end{align}\\
\addtocounter{equation}{1}
    \DivC\Vb \label{eq:H_konti}
        = 0
\end{gather}
\end{subequations}
holds for $ \dot{\rho} = 0 $ and given initial conditions for $ \Vb $ and mass density $ \rho\in\tangentScal $. \\
\end{model}

For $\meanc_0 = 0$ \eqref{eq:H_R3}
also follows by assuming $ \beta = \frac{2}{3} $ in \eqref{eq:BE_R3}. Accordingly, the viscosity rescales as $ \tilde{\coeffIF} = \coeffIF (1+\frac{\xi}{3})^2 $, 
although we may also simply set $ \xi=0 $ \oeda, since the anisotropy coefficient no longer appears anywhere else, and thus no actual anisotropic viscosity contributes to the solution.
As expected, all Gaussian curvature-elastic terms vanish independently of $ \bar{\kappa} $.
It should be noted that if the assumption $ \beta=\frac{2}{3} $ is justified by the fact that the immobility is small, \ie\ $ M\approx 0 $, such that the molecular equation \eqref{eq:full_R3_mol} relaxes on a much faster timescale than the fluid equation \eqref{eq:full_R3_fluid}, then the immobility term in the fluid equation can obviously be neglected.
In this case, the material ($ \Phi=\mfrak $) and Jaumann ($ \Phi=\jau $) models are indistinguishable.
Within the covariant differential calculus and the tangential-normal splitting of the fluid equations, the governing equations take the form:
\begin{subequations}\label{eq:H_cov}
\begin{gather}
    \begin{align}
        \rho\ab 
            &= \nabla\bp
                + \tilde{\coeffIF} \left( \Delta\vb + \gaussc\vb - 2\left( \shop - \frac{\meanc}{2}\IdS \right)\nabla\vnor \right)\\
            &\hspace{3em} + 2M\delta_{\mfrak}^{\Phi}\left( \gaussc\vb - \shop\left( \nabla\vnor + \meanc\vb \right) \right)\notag\\
        \rho\anor
            &= \meanc\bp
                + 2\tilde{\coeffIF}\left( \shop\dbdot\nabla\vb - \vnor\left( \meanc^2 - \gaussc \right) \right)
                + 2M\delta_{\mfrak}^{\Phi}\left( \Delta\vnor + \shop\dbdot\nabla\vb + \nabla_{\vb}\meanc \right)\\
            &\hspace{3em}
                -\kappa\left( \Delta\meanc + \frac{1}{2}\left( \meanc-\meanc_0 \right)\left( \meanc \left( \meanc+\meanc_0 \right)- 4\gaussc \right) \right) \notag
    \end{align}\\
    \div\vb
        = \meanc\vnor
\end{gather}
\end{subequations}
The associated Jaumann-Model ($ \Phi=\jau $), \resp\ neglected immobility ($ M=0 $), is consistent
with \citet{ReutherNitschkeEtAl_JoFM_2020, KrauseVoigt_JoCP_2023, Olshanskii_PoF_2023, SischkaNitschkeEtAl_FD_2025,Sauer_JoFM_2025}, 
and with \citet{ArroyoDeSimone_PRE_2009, Torres-SanchezMillanEtAl_JoFM_2019, ZhuSaintillanEtAl_JoFM_2025} in the Stokes limit.
We would like to emphasize once again that the material model ($ \Phi=\mfrak $, $ M > 0 $) is, also in this setting, not invariant under rigid-body rotations.
This lack of invariance may in fact be physically meaningful if a kind of rotational damping of the lipid molecules is assumed. 

While the recovery of these known surface (Navier-)Stokes-Helfrich models for fluid deformable surfaces is appealing, their derivation differs. \citet{BrandnerReuskenEtAl_IFB_2022} summarizes different derivations of the surface Navier-Stokes equations. Their extensions towards \eqref{eq:H_R3} or \eqref{eq:H_cov} in \citet{ArroyoDeSimone_PRE_2009, Torres-SanchezMillanEtAl_JoFM_2019,ReutherNitschkeEtAl_JoFM_2020} always consider an ad hoc additional Helfrich energy. The derivations in Section \ref{sec:derivations} do not require such an approach, the Helfrich-terms naturally follow from the underlying surface liquid crystal models, similarly to the original derivation of \citet{Helfrich_ZfNA_1971}. 


\section{Derivations} \label{sec:derivations}

\subsection{Q-Tensor ansatz}\label{sec:QAnsatz}
Essentially, we adopt a similar approach to that of \citet{Helfrich_ZfNC_1973}.
However, instead of considering a polar field normal to the surface $ \surf $ to describe a lipid bilayer,
we employ an apolar field represented by an uniaxial Q-tensor field $ \Qb\in\tangentQR $  with a normal eigenvector:
\begin{align}\label{eq:QAnsatz}
    \Qb
        &= S\left( \normal\otimes\normal - \frac{1}{3}\Id \right)
         = \beta\left( \normal\otimes\normal - \frac{1}{2}\IdS \right) \formComma
\end{align}
where $\Id = \IdS + \normal\otimes\normal$ is valid and $ S=\frac{3}{2}\beta\in\tangentScal $ is the scalar order field, \resp\ $ \beta=\frac{2}{3} S\in\tangentScal $ the eigenvalue field to the eigenvector field $ \normal $,
\ie\ it holds $ \Qb\normal=\beta\normal $.
With this ansatz, the lipid molecules are, on average, oriented orthogonally to the surface tangential planes and the amount of order is given by the scalar order $S$
as weighted average of the molecular orientation \citep{MottramNewton_a_2014} deviating from $\normal$.
For convenience, we use $\beta$ instead of $S$ in this paper.
Note that $\Qb$ in \eqref{eq:QAnsatz} is surface conforming without any flat degenerated part, see \citet{NitschkeVoigt_AiDE_2025}.

The advantage of modeling lipid layers as Q-tensor fields lies in the ability to describe molecular alignment as being, on average, normal to the surface, while also incorporating an additional degree of freedom for the statistical order. 
This approach allows us to build on established surface Beris--Edwards models to develop hydrodynamic descriptions of lipid bilayers, as discussed in \cref{sec:monolayer}.
A limitation of \eqref{eq:QAnsatz}, however, is that it is an apolar ansatz, \ie\ due to head-tail symmetry of the molecules we can only represent symmetric lipid bilayers a priori.
The symmetry break required to obtain asymmetric lipid bilayers must occur elsewhere in the process.
This is addressed in \cref{sec:bilayer}. 

\subsection{Surface Beris--Edwards model for symmetric lipid bilayers}\label{sec:monolayer}

In this section we use the surface conforming Beris--Edwards models in \citet{NitschkeVoigt_AiDE_2025} as a starting point and incorporate a suitable constraint providing the ansatz \eqref{eq:QAnsatz}.
Since ansatz \eqref{eq:QAnsatz} as well as the surface conforming models comprise up-down symmetry, \ie\ the exterior and interior of the surface are indistinguishable in any way, we could only describe symmetric lipid bilayers, see \cref{fig:sym_bilayers}, without introducing any additional symmetry-breaking mechanism.
For the reader’s convenience, a summary of the surface Beris--Edwards model from \citet{NitschkeVoigt_AiDE_2025} together with the active terms introduced in \citet{NitschkeVoigt_PotRSAMPaES_2025} is given in \cref{sec:full_surface_conforming_models}.

A reasonable orthogonal decomposition of $ \Qb\in\tangentQR $ is provided by
\begin{align}\label{eq:QDecomposition}
    \Qb = \qb + \etab\otimes\normal + \normal\otimes\etab + \beta\left( \normal\otimes\normal - \frac{1}{2}\IdS \right)\formComma
\end{align}
where $ \qb\in\tangentQS $ yields the flat degenerated part, $ \etab\in\tangentS $ the surface non-conforming part, and $ \beta\in\tangentScal $ the uniaxial normal part.
The surface conforming constraint $ \etab=\nullb $ is already incorporated in the surface conforming Beris--Edwards model \eqref{eq:model_conforming}.
Therefore, we only need to consider a ``No flat Degeneracy'' ($ \qb = \nullb $) constraint for our purpose. We account for possible constraint forces in \cref{sec:NoFlatDegeneracy} and conclude that it suffices to omit the flat degenerated molecular equation \eqref{eq:model_conforming_qtensor} and simply substitute $\qb=\nullb$ throughout the remaining model.

We first address the elastic contributions arising from the elastic energy functional
\begin{align}\label{eq:EL_energy}
    \energyEL 
       &= \frac{L}{2}\normHsq{\tangentR[^3]}{\nablaC\Qb}
\end{align}
The relevant terms in \cref{tab:forces_conforming} yield
\begin{align*}
    \sigmabEL
        &= -\frac{3}{2}L\left( \nabla\beta\otimes\nabla\beta - \frac{\normsq{}{\nabla\beta}}{2}\IdS 
                                +  3\meanc\beta^2\left( \shop - \frac{\meanc}{2}\IdS\right) \right)\formComma\\
    \zetabEL
        &= -\frac{3}{2}L\left( 2\shop\nabla\beta + \beta\nabla\meanc \right)\formComma\\
    \omegaEL
        &= L\left( \Delta\beta  - 3\beta \left( \meanc^2 - 2\gaussc \right) \right)\formPeriod
\end{align*}
Note that the tangential elastic fluid force $ \div\sigmabEL $ 
combined with the elastic part of the tangential surface conforming constraint force $ -3\beta\shop\zetabEL $ results in
\begin{align*}
    \div\sigmabEL - 3\beta\shop\zetabEL \hspace{-7em}\\
        &= -\frac{3}{2}L \Bigg( \Delta\beta\nabla\beta + \nabla^2\beta \nabla\beta - \nabla^2\beta \nabla\beta 
                                + 3\beta^2\left( \shop\nabla\meanc - \frac{\meanc}{2}\nabla\meanc  \right)
                                                + 6\meanc\beta\left( \shop\nabla\beta - \frac{\meanc}{2}\nabla\beta  \right) \\
         &\hspace{4em}   + \frac{3}{2}\meanc\beta^2\nabla\meanc     
                         -3\beta\left( 2\meanc\shop\nabla\beta - 2\gaussc\nabla\beta + \beta\shop\nabla\meanc \right)\Bigg)\\
         &= -\frac{3}{2}L \left( \Delta\beta  - 3\beta \left( \meanc^2 - 2\gaussc \right) \right) \nabla\beta
          = -\frac{3}{2} \omegaEL \nabla\beta \formComma
\end{align*}
since $ \div\shop=\nabla\meanc $ and $ \shop^2=\meanc\shop - \gaussc\IdS $ is valid.
The normal elastic fluid force $ \shop\dbdot\sigmabEL $ 
combined with the elastic part of the normal surface conforming constraint force $ 3\div(\beta\zetabEL) $ becomes
\begin{align*}
    \shop\dbdot\sigmabEL + 3\div(\beta\zetabEL)
        &= -\frac{3}{4}L\Big( 14\nabla\beta\shop\nabla\beta - \meanc\normsq{}{\nabla\beta} \\
         &\hspace{2em} +3\beta\left( 4\shop\dbdot\nabla^2\beta + 8\nabla\meanc\nabla\beta 
                                                +\beta\left( 2\Delta\meanc + \meanc(\meanc^2 - 4\gaussc) \right)\right)\Big) \formPeriod
\end{align*}
This implies that elasticity of symmetric lipid bilayers already comprises the bending force $ \fnorBE\normal $ for surfaces with up-down symmetry and mean curvature-elastic moduli $ \kappa = \frac{9}{2}L\beta^2$. Since the lipid molecules already define the surface geometry, we omit the lipid-independent bending force $\fnorBE$ and retain only the one arising from the elasticity in terms of the scalar order parameter. 
The molecular thermotropic force results in
\begin{align*}
    \omegaTH
        &= -\left(2a + b\beta + 3c\beta^2\right)\beta \formPeriod
\end{align*}
Other thermotropic contributions have no influence on the model.
The immobility terms yield
\begin{align*}
    \sigmabIM &= \nullb\formComma
    &\zetabIM &= \begin{cases}
                        \nullb 
                             &\text{, for }  \Phi = \jau \formComma\\
                        \frac{3}{2}M \beta \left(\nabla\vnor + \shop\vb\right)
                                &\text{, for } \Phi = \mfrak \formComma
                 \end{cases}
\end{align*}
see \eqref{eq:IM_fluid_force} for the corresponding fluid forces.
Nematic viscosity terms become
\begin{align*}
    \sigmabNV^0 + \xi\sigmabNV^1 + \xi^2\sigmabNV^2
        &= \tsigmabNV + \frac{\coeffIF\xi}{2}\dot{\beta}\left( 1 + \frac{\xi\beta}{2} \right)\IdS \formComma
    &\widetilde{\omega}^{2}_{\NV}
        &= 0 \formPeriod
\end{align*}
Due to inextensibility, we only have to consider
\begin{align*}
    \tsigmabNV
        &= \coeffIF\left( 1 - \frac{\xi\beta}{2} \right)^2 \left( \nabla\vb + \nabla^T\vb - 2\vnor\shop \right) \formComma
\end{align*}
see \eqref{eq:NV_fluid_force} for the corresponding fluid forces.
Nematic activity does not contribute in a inextensible media, \ie
\begin{align*}
    \sigmabNA &= \nullb \formPeriod
\end{align*}
Nevertheless, even though it does not affect the inextensible model, we note the presence of an active pressure 
$p_{\AC} = \alpha_{\IA} - \frac{\alpha_{\NA}}{2}\beta$, comprising a nematic part controlled by parameter $\alpha_{\NA}\in\R$ and a geometric part controlled by the parameter $\alpha_{\IA}\in\R$, see \citet{NitschkeVoigt_PotRSAMPaES_2025}. 
The corresponding pressure force therefore reads
\begin{align*}
    \FbAC
        &= \DivC\left( \left( \alpha_{\IA} - \frac{\alpha_{\NA}}{2}\beta \right)\IdS \right)
         = \alpha_{\IA}\GradC 1 - \frac{\alpha_{\NA}}{2} \GradC\beta\\
        &= - \frac{\alpha_{\NA}}{2}\nabla\beta + \left( \alpha_{\IA} - \frac{\alpha_{\NA}}{2}\beta \right)\meanc\normal \formPeriod
\end{align*}
Eventually, the surface Beris--Edwards model \eqref{eq:model_conforming}, the ``No flat Degeneracy'' constraint, and rewriting all relevant terms in the $ \R^3 $-calculus, yield the model \eqref{eq:BE_R3} for symmetric lipid bilayers.

\subsection{Hydrodynamic surface Landau--Helfrich model for asymmetric lipid bilayers}\label{sec:bilayer}

In this section, we derive the surface Landau--Helfrich energy as a special case of a polarized surface Landau--De Gennes energy, which consists of a nematic elastic contribution, a thermotropic contribution, similar to \citet{NitschkeNestlerEtAl_PotRSAMPaES_2018,NitschkeReutherEtAl_PRSA_2020}, and an up-down symmetry-breaking mechanism. 
The reduction to the lipid-molecule level is carried out using the ansatz \eqref{eq:QAnsatz}.
For this purpose, we identify all invariants for asymmetric lipid bilayers up to a certain order,
similar to the procedure in \citet{Helfrich_ZfNC_1973} and \citet{Alexe-Ionescu_PLA_1993}.
We then reformulate these into a form analogous to the Helfrich energy, which generalizes it accordingly.
The governing equations derived from this formulation and the Lagrange--D’Alembert principle for moving surfaces \citep{NitschkeVoigt_AiDE_2025} are provided in \cref{sec:variations}.

The componentwise surface derivative of \eqref{eq:QAnsatz} results in
\begin{align}\label{eq:nablaCQAnsatz}
    \nablaC\Qb 
        &= \left( \normal\otimes\normal - \frac{1}{2}\IdS \right) \otimes\nabla\beta
            -\frac{3}{2}\beta\left( \normal\otimes\shop + (\normal\otimes\shop)^{T_{(1\, 2)}}\right) \formComma
\end{align}
where $ [(\normal\otimes\shop)^{T_{(1\, 2)}}]_{ABC} = \normalC_{B}\shopC_{AC} $ in arbitrary Euclidean coordinates.
Both summands and all their transpositions are mutual orthogonal or symmetric, 
\ie\ all scalar valued quadratic invariants of \eqref{eq:nablaCQAnsatz} are given by 
$ \normsq{\tangentS}{\nabla\beta} $,
and $ \normsq{\tangentS[^2]}{\beta\shop} = (\meanc^2-2\gaussc)\beta^2 $
or $ (\Tr\shop)^2 \beta^2 = \meanc^2 \beta^2 $.
Therefore, we could reduce the quadratic energy approach 
\begin{align*}
    \int_{\surf} l_{A_1,A_2,A_3,B_1,B_2,B_3}[\nablaC\Qb]^{A_1,A_2,A_3}[\nablaC\Qb]^{B_1,B_2,B_3} \dS
\end{align*}
to the elastic energy potential
\begin{align}\label{eq:energytel}
    \energytel
        := \frac{l_1}{2} \normHsq{\tangentS}{\nabla\beta} + \frac{l_0}{2}\normHsq{\tangentScal}{\meanc\beta} - \bar{l}_0\int_{\surf}\gaussc\beta^2\dS
\end{align}
with  material parameters $ l_1 \ge 0 $ and $ l_0 \ge \bar{l}_0 \ge 0 $.
For instance, if we substitute ansatz \eqref{eq:QAnsatz} into the one-constant elastic energy potential \eqref{eq:EL_energy}, we obtain 
\begin{align*}
    \energyEL 
        &= \frac{3}{4}L\left( \normHsq{\tangentS}{\nabla\beta} + 3\normHsq{\tangentScal}{\meanc\beta} - 6\int_{\surf}\gaussc\beta^2\dS \right)\formComma
\end{align*}
\ie\ the parameter relation would be $ l_1 = \frac{3L}{2} $ and $ l_0= \bar{l}_0 = \frac{9L}{2} $.
Other conceivable elastic energy potentials that are quadratic in $ \Qb $
--- for instance, those involving the commonly used elastic parameters $ L_1=L, L_2, L_3 $ \citep{MoriJrKellyBos_JJoAP_1999, MottramNewton_a_2014} ---
yield energies with different 
proportions\footnote{If we consider a constant extension in normal direction of $ \Qb $, \cf\ \citet{NitschkeNestlerEtAl_PotRSAMPaES_2018},
                        we obtain 
                        $ \frac{L_2}{2}\normHsq{\tangentR}{\DivC\Qb} = \frac{L_2}{8}\left( \normHsq{\tangentS}{\nabla\beta} + 9\normHsq{\tangentScal}{\beta\meanc} \right)$
                        and
                        $ \frac{L_3}{2}\innerH{\tangentR[^3]}{\nablaC\Qb,\nablaC^T\Qb} = \frac{L_3}{8}\left( \normHsq{\tangentS}{\nabla\beta} + 9\normHsq{\tangentScal}{\beta\meanc} - 18\int_{\surf}\gaussc\dS \right)$
                        for instance.}.
Note that \eqref{eq:energytel} states Helfrich's bending energy \citep{Helfrich_ZfNC_1973} for constant $ \beta $ and neglected spontaneous curvature, 
where the curvature-elastic moduli yield $ \kappa= l_0\beta^2 $ and $ \bar{\kappa} = \bar{l}_0\beta^2 $.

In order to incorporate also spontaneous curvature effects we have to break the up-down symmetry of our approach.
The elastic potential \eqref{eq:energytel} cannot accomplish this on its own. 
In the field of liquid crystals, such a polarization has already been derived via the piezoelectric effect \citep{Meyer_PRL_1969,Helfrich_ZfNA_1971}. 
Essentially, this involves assuming an asymmetry in the physics of the nematic phase, which exhibits either splay and/or bend structures. 
Lipid bilayers, with average molecule directions perpendicular to the surface, may likewise exhibit a splay, since parallel alignment cannot be sustained on a curved surface, \cf\ \cref{fig:asym_bilayers}, 
but no bending asymmetry, which would require molecules to possess, on average, a tangential component.
Such a splay asymmetry can originate from a variety of mechanisms \citep{ZimmerbergKozlov_NRMCB_2005},
for instance, if the inner layer consists of a different molecular composition than the outer one \citep{London_AoCR_2019,LorentLeventalEtAl_NCB_2020,BogdanovPyrshevEtAl_SA_2020,EnokiHeberle_PotNAoS_2023}, 
or if the molecular density differs between the two sides \citep{RozyckiLipowsky_TJoCP_2015,SreekumariLipowsky_TJoCP_2018,GhoshSatarifardEtAl_NL_2019,SreekumariLipowsky_SM_2022,LipowskyGhoshEtAl_B_2023}, 
or if binding of proteins, such as coat or scaffolds proteins, is included \citep{HuZhangEtAl_PRE_2007,Kirchhausen_NCB_2012,SaleemMorlotEtAl_NC_2015,KahramanLangenEtAl_SR_2018,MironovMironovEtAl_IJoMS_2020}. 
This does not violate the apolarity of the average lipid orientations in normal direction, but rather pertains to the physical conditions that a bilayer might inherently possess.
In \citep{GoversVertogen_PRA_1984,BarberoDozovPalierneDurand_PRL_1986, Alexe-Ionescu_PLA_1993, MottramNewton_a_2014} flexoelectric polarization in terms of Q-tensors is already discussed.
Considering the normal direction $ \normal $ as polarization direction yields 
\begin{align*}
    \normal\DivC\Qb
        &= -\frac{3}{2}\meanc\beta \formComma
    &\normal(\nablaC\Qb)\dbdot\Qb
        &= \frac{3}{4}\meanc\beta^2 \formComma
    &\normal\Qb\DivC\Qb
        &= -\frac{3}{2}\meanc\beta^2 \formComma
\end{align*}
for the non-vanishing ($ E_1,E_5,E_6 $) terms in \citet{Alexe-Ionescu_PLA_1993}.
As we see below, the thermotropic potential is a fourth-order polynomial in $ \beta $ and $ \meanc $, albeit with vanishing coefficients for the mean curvature. 
Therefore, we extend the expansion of the polarization up to third order,
which, additionally, comprises the non-vanishing 
terms\footnote{In notation and terms of \citet{Alexe-Ionescu_PLA_1993}, the flexoelectric coefficients tensor $ E_{i\alpha\beta\gamma} $,
                \soth\ $ \normalC_i P_i = \normalC_i E_{i\alpha\beta\gamma} Q_{\alpha\beta,\gamma} $, is extended to
                $
                    E_{i\alpha\beta\gamma}
                        = \ldots 
                            + \frac{1}{2}E_7\left(Q_{i\alpha}Q_{\beta\gamma} + Q_{i\beta}Q_{\alpha\gamma}\right)
                            + E_8 Q_{i\gamma}Q_{\alpha\beta}
                            + \frac{1}{2}E_9 Q_{\delta\mu}Q_{\delta\mu}\left( \delta_{i\alpha}\delta_{\beta\gamma} + \delta_{i\beta}\delta_{\alpha\gamma} \right)
                            + E_{10} Q_{\delta\mu}Q_{\delta\mu} \delta_{i\gamma}\delta_{\alpha\beta} 
                            + E_{11} \delta_{i\gamma}Q_{\alpha\delta}Q_{\beta\delta}
                            + E_{12} \delta_{\alpha\beta}Q_{i\delta} Q_{\gamma\delta}
                            + \frac{1}{2}E_{13} Q_{\gamma\delta}\left( \delta_{i\alpha}Q_{\beta\delta} + \delta_{i\beta}Q_{\alpha\delta} \right)
                            + \frac{1}{2}E_{14} Q_{i\delta}\left( \delta_{\alpha\gamma}Q_{\beta\delta} + \delta_{\beta\gamma}Q_{\alpha\delta}\right)\formPeriod
                $
                Due to polarization direction $ \normal $, which is a eigenvector of $ \Qb $, and $ (\nablaC\Qb)\normal=\nullb $, the $ E_8 $-, $ E_{10} $-, $ E_{11} $-, and $ E_{12} $-term vanish.
                }
\begin{align*}
    \normal\Qb(\nablaC\Qb)\dbdot\Qb
            &= \frac{3}{4}\meanc\beta^3\formComma
    &(\Tr\Qb^2)\normal\DivC\Qb
            &= -\frac{9}{4}\meanc\beta^3\formComma \\
    \normal(\nablaC\Qb)\dbdot\Qb^2
            &= -\frac{3}{8}\meanc\beta^3\formComma
    &\normal\Qb^2\DivC\Qb
            &= -\frac{3}{2}\meanc\beta^3\formPeriod
\end{align*}
Therefore, the polarized elastic energy, up to forth polynomial order, \resp\ third order in $ \beta $, is of the form 
\begin{align*}
    \int_{\surf}  \left(k_1 + \frac{k_2}{2}\beta + \frac{k_3}{3}\beta^2\right)\meanc\beta\dS
\end{align*}
where $ k_1,k_2,k_3\in\R $ are the polarization parameters.
Order polarization in terms of $\nabla\beta$ is not included as a consequence.
Eventually, we approach the polarized elastic energy by
\begin{align*}\label{eq:energyel}
	\energyel
    	&:= \frac{l_1}{2} \normHsq{\tangentS}{\nabla\beta}
    		+ \int_{\surf} \frac{l_0}{2} \meanc^2\beta^2 - \bar{l}_0 \gaussc \beta^2 +  \frac{k_3}{3}\meanc\beta^3 + \frac{k_2}{2} \meanc\beta^2 + k_1\meanc\beta \dS    \formPeriod      
\end{align*} 
Using \eqref{eq:QAnsatz} the thermotropic energy becomes
\begin{align*}
    \energyTH
    	&= \int_{\surf} a \Tr\Qb^2 + \frac{2b}{3}\Tr\Qb^3 + c\Tr\Qb^4 \dS\\
        &= \int_{\surf} \frac{3a}{2} \beta^2 + \frac{b}{2}\beta^3 +\frac{9c}{8}\beta^4\dS
         = \int_{\surf} \frac{A}{2} \beta^2 + \frac{B}{3}\beta^3 + \frac{C}{4}\beta^4\dS\formComma
\end{align*}
\ie\ $ A=3a $, $ B = \frac{3b}{2} $, and $ C=\frac{9c}{2} $ holds for the thermotropic coefficients.
Due to the zeroth-order derivative terms, the elastic and thermotropic energy should not be considered separately.
The polarized surface Landau--de Gennes energy now reads 
\begin{align}
    \energyLdG
    	&:= \energyel + \energyTH\notag\\
    	 &= \frac{l_1}{2} \normHsq{\tangentS}{\nabla\beta} \label{eq:LdG_polarized}\\
    	 &\quad + \int_{\surf} k_1\meanc\beta + \frac{A + k_2 \meanc + l_0 \meanc^2 - 2\bar{l}_0\gaussc}{2} \beta^2 
                            + \frac{B+k_3\meanc}{3}\beta^3 + \frac{C}{4}\beta^4 \dS\formPeriod \notag
\end{align}
Note that this potential has a minimum for $ l_1 \ge 0$, $ l_0 \ge \bar{l}_0 \ge 0 $, $ C>0 $ and $ k_3 \in \left( -\frac{3}{2}\sqrt{C \left(2l_0 - \bar{l}_0\right)}, \frac{3}{2}\sqrt{C \left(2l_0 - \bar{l}_0\right)} \right) $ \eqref{eq:k3_ival}, see \cref{app:material_parameter_constrains}.

\begin{figure}[t]
	\centering
	\hfill
	\includegraphics[width=0.48\textwidth]{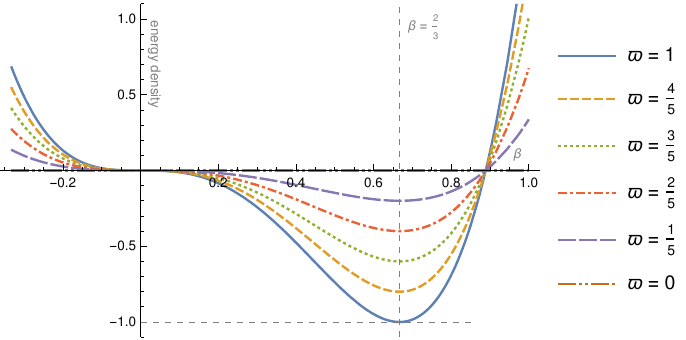}
	\hfill
	\includegraphics[width=0.48\textwidth]{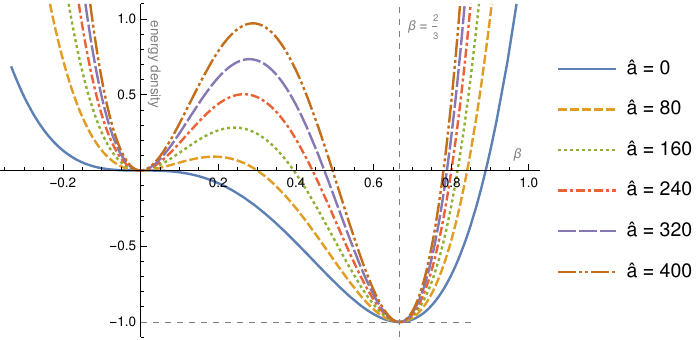}
	\hfill
	\caption{Energy density plots of the double-well potential \eqref{eq:DW}.
            In the left plot we stipulate $ \hat{a}=0 $ and choose some values for $ \varpi $. 
            Their additive inverse equal the minimum values at the fully ordered state $ \beta=\frac{2}{3} $.
            In the right plot we use $ \varpi=1 $ and specify some values for $ \hat{a} $. 
            Only $ \hat{a}=0 $ yields an inflection point at the isotropic state $ \beta=0 $ instead of a minimum for $ \hat{a} > 0 $.}
	\label{fig:DW}
\end{figure}

\begin{figure}[t]
	\centering
	\hfill
	\includegraphics[width=0.48\textwidth]{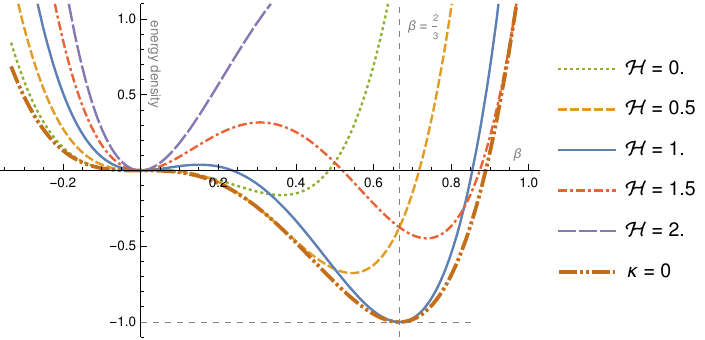}
	\hfill
	\includegraphics[width=0.48\textwidth]{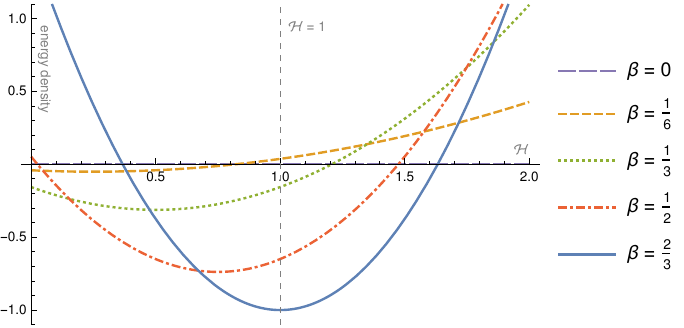}
	\hfill
	\caption{Energy density plots of the energy \eqref{eq:LdGReduced__hatH0_equals_0} ($\hat{\meanc}_0=\hat{a}=0$) with $L=\bar{\kappa}=0$.
				We stipulate $\kappa=5$ and $\meanc_{0}=\varpi=1$.
				In the left plot we fix the mean curvature $\meanc$ for some fixed values around $\meanc_{0}=1$.
				Additional we add the $\kappa=0$ case, which does not depend on $\meanc$ and leads $\beta$ to the fully ordered state $ \beta=\frac{2}{3}$.
				The right plot shows the dependency \wrt\ $ \meanc $ for some fixed $\beta$'s.}
	\label{fig:LdGInBetaAndH}
\end{figure}

The material parameters in \eqref{eq:LdG_polarized} appear to be somewhat unintuitive, as they represent purely formal expansion parameters. 
Therefore, we aim to reformulate \eqref{eq:LdG_polarized} in the form 
\begin{align*}
    \energyLH
        	&:= \frac{3L}{4} \normHsq{\tangentS}{\nabla\beta}
        		+ \int_{\surf} \frac{f_{\kappa}(\beta)}{2}\left( \meanc - f_{\meanc_{0},\hat{\meanc}_{0}}(\beta) \right)^2 
        						-f_{\bar{\kappa}}(\beta)\gaussc
        						+ f_{\varpi,\hat{a}}(\beta)
        			\dS \formComma
\end{align*}
such that $f_{\kappa}$, $f_{\meanc_{0},\hat{\meanc}_{0}}$, $f_{\bar{\kappa}}$, and $f_{\varpi,\hat{a}} $, are polynomials in $\beta$, 
\eqref{eq:LdG_polarized} is satisfied as generally as possible, 
and $f_{\kappa}\left( \frac{2}{3} \right) = \kappa$,
$f_{\meanc_{0},\hat{\meanc}_{0}}\left( \frac{2}{3} \right) = \meanc_{0}$,
$f_{\bar{\kappa}}\left( \frac{2}{3} \right) = \bar{\kappa} $,
and $f_{\varpi,\hat{a}}\left( \frac{2}{3} \right) = -\varpi $ (local minimum)
hold at the fully ordered state. 
By providing a sufficient polynomial ansatz, incorporating the required conditions, and comparing coefficients, we obtain the quadratic curvature-elastic polynomials
$ f_{\kappa}(\beta) = \frac{9}{4}\kappa\beta^2 $
and $f_{\bar{\kappa}}(\beta) = \frac{9}{4}\bar{\kappa}\beta^2 $,
linear spontaneous curvature polynomial
$ f_{\meanc_{0},\hat{\meanc}_{0}}(\beta) = \hat{\meanc}_{0} + \frac{3}{2} ( \meanc_{0} - \hat{\meanc}_{0} )\beta $,
and double-well potential
\begin{align}\label{eq:DW}
    f_{\varpi,\hat{a}}(\beta) 
        &= \frac{9}{4}\beta^2\left( \frac{\hat{a}}{9}\left( \frac{2}{3}-\beta \right)^2 
            - \frac{3\varpi}{2}\beta\left( 4-\frac{9}{2}\beta \right) \right) \formComma
\end{align}
see \cref{fig:DW}. Similar linear spontaneous curvature polynomials have been considered to interpolate between spontaneous curvature values for coexisting, e.g. liquid ordered and liquid disordered, phases \citep{Seifert_PRL_1993,Taniguchi_PRL_1996,
BachiniKrauseEtAl_JoFM_2023,SischkaEtAl_CMAME_2025}. 
As a consequence, the polarized surface Landau--de Gennes energy becomes 
\begin{align}\label{eq:LH}
    \energyLH
    	&= \frac{3}{4} L \normHsq{\tangentS}{\nabla\beta} \\
        &\quad+ \int_{\surf} \frac{9}{4}\beta^2\left( \frac{\kappa}{2} \left( \meanc - \left(\hat{\meanc}_{0} + \frac{3}{2}\left( \meanc_{0} - \hat{\meanc}_{0} \right)\beta\right) \right)^2 
                        - \bar{\kappa}\gaussc 
							\right) 
							+ f_{\varpi,\hat{a}}(\beta)\ \dS\formComma\notag
\end{align}
and we name it surface Landau--Helfrich energy.
Alternative approaches beyond this resulting linear interpolation of the spontaneous curvature with respect to a scalar order are conceivable; see, e.g., \citet{BartelsDolzmannEtAl_IaFBMACaA_2012}.
Note that the surface Landau--Helfrich energy \eqref{eq:LH} involves only 7 material parameters (listed in \cref{tab:material_parameter}),
instead of 9 as in the polarized surface Landau--de Gennes energy \eqref{eq:LdG_polarized}.
Its material parameters are uniquely determined as follows:
\begin{align}\label{eq:material_parameter_relation}
\begin{aligned}
    A 
       	&= \frac{9\kappa}{4}\hat{\meanc}_{0}^2 + \frac{2}{9}\hat{a} \formComma
    & k_1
    	&= 0 \formComma
    & l_0
        &= \frac{9\kappa}{4} \formComma \\
    B
       	&= \frac{81\kappa}{8}  \left( \meanc_{0} - \hat{\meanc}_{0} \right)\hat{\meanc}_{0} - \frac{2\hat{a} + 81\varpi}{2} \formComma
    & k_2
    	&= -\frac{9\kappa}{2} \hat{\meanc}_{0} \formComma
    & \bar{l}_0
        &= \frac{9\bar{\kappa}}{4} \formComma\\
    C
       	&= \frac{81\kappa}{8}  \left( \meanc_{0} - \hat{\meanc}_{0} \right)^2 + \frac{ 4\hat{a} + 243\varpi}{4} \formComma
    & k_3
       	&= -\frac{81\kappa}{8}  \left( \meanc_{0} - \hat{\meanc}_{0} \right) \formComma
    & l_1
        &= \frac{3 L }{2}\formPeriod
\end{aligned}
\end{align}
The restriction of $ k_3 $ yields now the restriction \eqref{eq:DW_parameter_regime} of the double-well parameter $ \hat{a} $ and $ \varpi $, see \cref{app:material_parameter_constrains}.
It should be noted that the case $C=0$ is included here, although not discussed in \cref{app:material_parameter_constrains}. 
Nevertheless, this yields $\meanc_{0} = \hat{\meanc}_{0}$, and $ \hat{a} = \varpi = 0  $, and thus
\begin{align}\label{eq:LdGReduced_H0_equals_hatH0}
    \energyLH\vert_{\meanc_{0} = \hat{\meanc}_{0}, \hat{a} = 0}  \hspace{-3em}&\\
            &= \frac{3}{4} L \normHsq{\tangentS}{\nabla\beta}
              + \int_{\surf} \frac{9}{4}\beta^2\left( \frac{\kappa}{2} \left( \meanc - \meanc_{0} \right)^2 
                                     - \bar{\kappa}\gaussc 
                                  - \frac{3\varpi}{2}\beta\left( 4-\frac{9}{2}\beta \right) \right) \dS\formComma \notag
\end{align}
which remains bounded below even for $\varpi = 0$.
The special case \eqref{eq:LdGReduced_H0_equals_hatH0} essentially states that the spontaneous curvature $\meanc_{0} = \hat{\meanc}_{0}$ does not depend on the scale order $\beta$, and that the isotropic state ($\beta\equiv0$) is possible but not stable due to $ \hat{a} = 0 $.
By contrast, another special case occurs when no spontaneous curvature $\hat{\meanc}_{0} = 0 $ is present in the isotropic state:
\begin{align}\label{eq:LdGReduced__hatH0_equals_0}
    \energyLH\vert_{\hat{\meanc}_{0}=0,\hat{a}=0} \hspace{-3em}&\\
        &= \frac{3}{4} L \normHsq{\tangentS}{\nabla\beta}
                    + \int_{\surf} \frac{9}{4}\beta^2\left( \frac{\kappa}{2} \left( \meanc - \frac{3}{2} \meanc_{0}\beta \right)^2 
                                - \bar{\kappa}\gaussc 
                                - \frac{3\varpi}{2}\beta\left( 4-\frac{9}{2}\beta \right) \right) \dS \formComma \notag
\end{align}
see \cref{fig:LdGInBetaAndH}.
It should be noted that $ \hat{\meanc}_{0} $ and $ \hat{a} $ become increasingly insignificant near the ordered state ($\beta\approx\frac{2}{3}$) anyway.
The converse of material parameter relation \eqref{eq:material_parameter_relation} is not uniquely determined, due to the reduction of the amount of parameters.
In \cref{tab:inverse_material_parameter_relation} we give an overview of these possible relations for the two special cases above. However, in order to address phase coexistence and account for the different properties of the liquid-ordered and liquid-disordered phases, as in \cite{BaumgartHessEtAl_N_2003}, these special cases are not sufficient and the full set of parameters, allowing for $\meanc_0 \neq \hat{\meanc}_0$ is required. 
\begin{table}
\centering
\begin{tabular}{lccc}
    assumptions:\hspace{2em}
	   & $\hat{\meanc}_0 = \meanc_0$
            &
			     & $\hat{\meanc}_0 = 0$ \\\clineX{2-4}
       &
            & $\hat{a}=0$ 
                &\\
\hline
	constraints:
		& $k_3 = 0$
            &
			     & $k_2 = 0$ \\
		& $2C + 3B = 0$
            &
			     & $A = 0$ \\
		& $k_2^2 = 4 l_0 A$
            &
			     & $4k_3^2 = 9 l_0 (2C + 3B)$\\\clineX{2-4}
        &
            &$k_1=0$
                &\\
\hline
	relations:
        & $\meanc_{0} = - \frac{k_2}{2 l_0}$
            &
        		& $\meanc_{0} = - \frac{2 k_3}{9 l_0}$\\\clineX{2-4}
        &
            &$L = \frac{2}{3}l_1$
                &\\
        &
		    & $\kappa = \frac{4}{9}l_0$
                &\\
        &
		    & $\bar{\kappa} = \frac{4}{9}\bar{l}_0$ 
                &\\
        &
		    & $\varpi = -\frac{2}{81} B$
                & 			
\end{tabular}
\caption{Inverse of material parameter relations \eqref{eq:material_parameter_relation} 
			for the special cases \eqref{eq:LdGReduced_H0_equals_hatH0} (left column) and \eqref{eq:LdGReduced__hatH0_equals_0} (right column).
            The middle column holds for both cases.
			Since the number of material parameters is reduced from 9 to 5, four constraints on the parameters follow.
			The relations are not unique due to some of these constraints.}
\label{tab:inverse_material_parameter_relation}
\end{table}
It should also be noted that the fully ordered regime ($ \beta = \frac{2}{3} $) yields the common Helfrich energy
\begin{align*}
    \energyLH\vert_{\beta\equiv\frac{2}{3}}
        &= \int_{\surf} \frac{\kappa}{2} \left( \meanc - \meanc_0 \right)^2 - \bar{\kappa}\gaussc - \varpi \dS
\end{align*}
with a constant offset ($-\varpi$), which has not any impact on variations for inextensible materials.

The derivation of the governing equations from the LH energy \eqref{eq:LH} and additional dynamical contributions, such as linear momentum and viscous forces, is carried out directly via the Lagrange--d’Alembert principle in \cref{sec:variations}.
\begin{figure}[t]
	\centering
	\hfill
	\includegraphics[width=\textwidth]{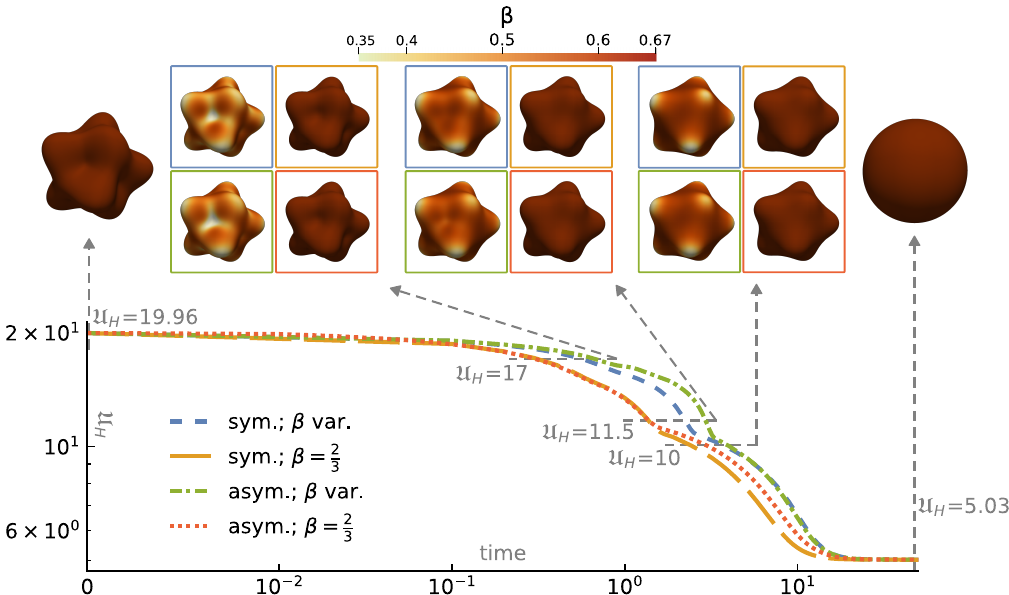}
	\hfill
	\caption{Evolution of the surface Beris--Edwards model (blue), the Navier--Stokes--Helfrich model with $\meanc_0 = 0$ (apricot), the hydrodynamic surface Landau--Helfrich model (green) and the Navier--Stokes--Helfrich model with $\meanc_0 = -1.77$ (red). The parameters used are stated in \cref{sec:Numerics}. Shown is a double logarithmic plot of the time evolution of the Helfrich energy $\mathfrak{U}_H = \int_\surf \frac{\kappa}{2}  \meanc^2\,d\surf$, with $\frac{\kappa}{2} = L = 0.1$, together with snapshots for different levels of $\mathfrak{U}_H$. $\mathfrak{U}_H=19.96$ corresponds to the initial condition, $\mathfrak{U}_H=5.03$ to the equilibrium configuration. In addition we show intermediate states for $\mathfrak{U}_H=17, 11.5$ and $10$ for each model. The frames of the images follow the same color coding as the plots. On the surface we show the value of the orientation field $\beta$ in color, ranging from $\beta=0.35$ in light yellow to $\beta=0.67$ in dark orange. While the shape correspond to the same Helfrich energy and therefor only slightly differ, their appearance in time differs between the models with slower shape evolution for variable $\beta$ far away from the equilibrium state.}
	\label{fig:results}
\end{figure}
\section{Discussion}
\label{sec:discussion}

We have derived various hydrodynamic liquid crystal models for lipid bilayers. They include a hydrodynamic surface Landau--Helfrich (LH) model for asymmetric lipid bilayers and a surface Beris--Edwards (BE) model for symmetric lipid bilayers. The latter directly follows from the surface conforming Beris--Edwards model derived in \citet{NitschkeVoigt_AiDE_2025} considering an additional "No flat Degeneracy" constraint. It is also a special case of the (LH) model if the polarization vanishes. The (LH) model builds on a special case of a surface polarized Landau--de Gennes energy and is derived using the Lagrange--d'Alembert principle. Both models consider a scalar order parameter that quantifies the degree of the molecular alignment along the surface normal. This order determines the bending properties, it influences the surface hydrodynamics, and, in case of the hydrodynamic surface Landau--Helfrich model, it introduces curvature-order coupling that breaks up-down symmetry. For the fully ordered state, the equations reduce to the surface \mbox{(Navier--)}Stokes--Helfrich model for fluid deformable surfaces \citep{ArroyoDeSimone_PRE_2009, Torres-SanchezMillanEtAl_JoFM_2019, ReutherNitschkeEtAl_JoFM_2020} and in its overdamped limit to the classical Helfrich model with local inextensibility constraint. The results therefore also provide an alternative derivation of the surface \mbox{(Navier--)}Stokes--Helfrich model or even the classical Helfrich model. However, the main contributions are the derived hydrodynamic liquid crystal models for lipid bilayers in \eqref{eq:BE_R3} and \eqref{eq:full_R3} for symmetric and asymmetric cases, respectively. In order to demonstrate the potential impact of the molecular order on the dynamics of these highly coupled systems we explore numerical simulations. Details on the numerical approach are provided in \cref{sec:Numerics}. In \cref{fig:results} the relaxation of a perturbed sphere is explored in a minimal setting. Already within this setting differences between the symmetric and asymmetric models with constant ($\beta = \frac{2}{3}$) and variable $\beta$ can be observed. However, the full potential of the proposed models requires detailed parameter studies and should be investigated in comparison with molecular dynamics and experimental studies. One direction to be explored is the coexistence of a liquid ordered and a liquid disordered state. This can be achieved by modifying the idealized parameters in the thermotropic energy $ \energyTH $. This will link the hydrodynamic surface Landau--Helfrich model for asymmetric lipid bilayers to models for two-phase systems of Jülicher-Lipowski type \citep{JuelicherEtAl_PRE_1996,BaumgartHessEtAl_N_2003,LowengrubEtAl_PRE_2009,ElliottEtAl_SIAMJAM_2010,BarrettEtAl_ESAIM_2017,BachiniEtAl_PoF_2023,BachiniKrauseEtAl_JoFM_2023,SischkaEtAl_CMAME_2025}. Phase coexistence between liquid ordered and liquid disordered phases still fall within the class of lipid bilayers, where the average lipid chain axis is approximately aligned with the membrane normal and orientational fluctuations are primarily captured by a scalar order parameter \citep{Seifert01021997,Marsh_2013}. This significantly simplifies the Q-tensor ansatz as the only remaining orientational degree of freedom is the scalar order parameter $ \beta $. Situations involving pronounced molecular tilt, hydrophobic mismatch, faceted gel-phase vesicles, or strong protein-induced director reorientation fall outside the scope of this simplified  formulation. Dealing with such situations requires the general surface Beris--Edwards model (\cite{NitschkeVoigt_AiDE_2025}) to be endowed with a polarization as considered in Section \ref{sec:bilayer} or at least retaining the surface non-conforming vector $ \etab\in\tangentS $ in \eqref{eq:QDecomposition} as a degree of freedom and describing the flat degenerate component $ \qb\in\tangentQS $ as a function of $ \etab $ and $ \beta $.
Solving the uniaxaility condition $ (\Tr\Qb^2)^3 - 6(\Tr\Qb^3)^2 = 0$ or $ \Qb^4-\frac{5}{6}(\Tr\Qb^2)\Qb^2+\frac{1}{9}(\Tr\Qb^2)^2\Id = \nullb $ \citep{NitschkeVoigt_AiDE_2025}
leads to
\begin{align*}
    \qb(\etab,\beta)
        &= \frac{1}{2}\left( \sqrt{9\beta^2 + 8\|\etab\|^2} - 3\beta \right)\left( \frac{\etab\otimes\etab}{\|\etab\|^2} - \frac{1}{2}\IdS \right) \formComma
\end{align*}
where $ \qb(\etab,\beta) \rightarrow \nullb $ holds for $ \etab\rightarrow\nullb $, which is the special case \eqref{eq:QAnsatz}.
Such an extension would already significantly increase both the modeling complexity and the effort required for a numerical approximation. We therefore refrain from pursuing this approach in the present paper. The same holds true for further model refinements, e.g. considering different leaflets or flip-flop mechanisms. 

\appendix

\section{Material parameter regime}\label{app:material_parameter_constrains}
The derivative part of the polarized surface Landau--de Gennes energy \eqref{eq:LdG_polarized} is bounded below simply by $l_1 = \frac{3L}{2} \ge 0$.
In contrast, the treatment of material parameters in the non-derivative part is not quite as straightforward.
For this purpose we define
\begin{align*}
	\int_{\surf} f \dS
    	&:= \energyLdG - \frac{l_1}{2}\normHsq{\tangentS}{\nabla\beta}\formPeriod
\end{align*}
Since the surface area is finite, we consider only the energy density $f$ in the following.
Due to dependencies between mean curvature $\meanc$ and Gaussian curvature $\gaussc$ we use the mutual independent principal curvatures $\kappa_1$ and $\kappa_2$ instead,
\ie\ we substitute $\meanc = \kappa_1 + \kappa_2$ and $\gaussc = \kappa_1\kappa_2$.
Therefore, $f$ is a polynomial of 4th order in $\kappa_1$, $\kappa_2$, and $\beta$:
\begin{align*}
	f 
		&= f_4 + f_3 + f_2 \formComma\\
	f_4(\kappa_1,\kappa_2,\beta)
		&= \frac{l_0}{2} \left( \kappa_1^2 + \kappa_2^2 \right)\beta^2
          +\left( l_0 - \bar{l}_0 \right)\kappa_1\kappa_2\beta^2
          +\frac{k_3}{3}\left( \kappa_1 + \kappa_2 \right)\beta^3
          + \frac{C}{4}\beta^4\formComma\\
    f_3(\kappa_1,\kappa_2,\beta)
    	&= \frac{k_2}{2} \left( \kappa_1 + \kappa_2 \right)\beta^2
    	   + \frac{B}{3}\beta^3\formComma\\
   	f_2(\kappa_1,\kappa_2,\beta)
   		&= k_1 \left( \kappa_1 + \kappa_2 \right)\beta + \frac{A}{2}\beta^2\formPeriod
\end{align*}
The asymptotic behavior is given by the leading polynomial $f_4$.
Since, a priori, we see no reasons to restrict the material parameter $0 \le \bar{l}_0 \le l_0$ and $ 0 < C $ any further,
we like to constrain the 3rd order polarizing parameter $k_3$ in the following.
The leading polynomial is evaluated in a spherical space in order to examine the asymptotic behavior: 
\begin{align*}
    f_4(\kappa_1,\kappa_2,\beta)
    	&= f_4(r\cos\theta\sin\varphi, r\sin\theta\sin\varphi, r\cos\varphi)
    	 = r^4 \frac{\cos^2\varphi}{12} s(\varphi,\theta)\formComma\\
   	s(\varphi,\theta)
   		&= 6\left(l_0 + 2\left(l_0 - \bar{l}_0\right)\sin\theta\cos\theta\right)\sin^2\varphi
   		  + 4 k_3 (\cos\theta + \sin\theta)\sin\varphi\cos\varphi
   		  + 3 C \cos^2\varphi\\
   		&=  6\left(l_0 + \left(l_0 - \bar{l}_0\right)\sin2\theta\right)\sin^2\varphi
  		  + 2\sqrt{2} k_3 \sin\left(\theta + \frac{\pi}{4}\right)\sin 2\varphi
  		  + 3 C \cos^2\varphi \formComma
\end{align*}
where $\varphi\in[0,\pi]$ and $\theta\in[0,2\pi)$,
\ie\ our argumentation is: 
\begin{align*}
    \forall\varphi,\theta:\quad s(\varphi,\theta) &>  0\quad\text{(to be shown)}
    &&\Rightarrow
    & \lim_{r\rightarrow\infty} f_4 &\in \{0, +\infty\}
    &&\Rightarrow
    &f &\text{ is bounded below.}
\end{align*}
Note that $ \lim_{r\rightarrow\infty} f_4 = 0 $ for $ s>0 $ only happens if $ \cos\varphi = 0 $ is valid, \ie\ $ \beta=0 $ and hence $ f=0 $.

The function $s$ has a unique minimum in $\varphi$, which can be computed by standard 
methods\footnote{For instance, find $ \varphi^*\in[0,\pi] $ \soth\ $\partial_\varphi s(\varphi^*,\theta) = 0$ and $ \partial^2_\varphi s(\varphi^*,\theta) > 0$ is valid.}.
Therefore, we can bound $s$ from below by
\begin{align*}
	s(\varphi,\theta)
		&\ge \min_{\varphi} s(\varphi,\theta)
    	= \frac{1}{2}\left( a_s(\theta) + b_s - \sqrt{(a_s(\theta) - b_s)^2 + c_s(\theta)^2} \right)\formComma
\end{align*}
where
\begin{align*}  	
    a_s(\theta)
    	&= 6\left(l_0 + \left(l_0 - \bar{l}_0\right)\sin2\theta\right)\formComma
   	&b_s &= 3 C \formComma
   	&c_s(\theta) 
   		&= 4\sqrt{2} k_3 \sin\left(\theta + \frac{\pi}{4}\right) \formPeriod
\end{align*}
Since $ a_s(\theta) \ge 0 $ and $C > 0$, we can square $a_s(\theta) + b_s$ for the inequality $s(\varphi,\theta) > 0$ and obtain the simpler problem
$ 4 a_s(\theta) b_s - c_s(\theta)^2 > 0 $.
With $ \sin^2\left(\theta + \frac{\pi}{4}\right) = \frac{1}{2}\left( 1 + \sin2\theta \right)$, this reads
\begin{align*}
    \left( 9 C \left(l_0 - \bar{l}_0\right) - 2 k_3^2 \right) \sin2\theta
    	+ 9 C l_0 - 2 k_3^2
    		&> 0 \formComma
\end{align*}
which must hold.
For $ k_3^2 \le \frac{9}{2} C \left(l_0 - \bar{l}_0\right) $ this inequality is always valid, since the minimum in $\theta$ on the left-handed side yields $ 9 C \bar{l}_0 $, which is positive by $C>0$ and $\bar{l}_0 \ge 0 $.
For $ k_3^2 > \frac{9}{2} C \left(l_0 - \bar{l}_0\right) $ the minimum on the left-handed side yields $ 9 C \left(2l_0 - \bar{l}_0\right) - 4 k_3^2 $.
This results in the upper bound $ k_3^2 < \frac{9}{4} C \left(2l_0 - \bar{l}_0\right) $, 
which is already less restrictive then the former $ k_3^2 \le \frac{9}{2} C \left(l_0 - \bar{l}_0\right) $.
Ultimately, we conclude that
\begin{align}\label{eq:k3_ival}
    k_3 &\in \left( -\frac{3}{2}\sqrt{C \left(2l_0 - \bar{l}_0\right)}, \frac{3}{2}\sqrt{C \left(2l_0 - \bar{l}_0\right)} \right)
\end{align}
is sufficient for $f$ to be bounded below.
Note that the parameter interval \eqref{eq:k3_ival} is even sharp for $ f $ to be bounded below, up to the edge cases $ k_3 = \pm\frac{3}{2}\sqrt{C \left(2l_0 - \bar{l}_0\right)} $,
which would imply that we would also need to examine the lower-order polynomial $ f_3 $, and possibly even $ f_2 $.
We shall not pursue these considerations here.

In the following, we align \eqref{eq:k3_ival} with the material parameters appearing in the surface Landau--Helfrich energy \eqref{eq:LH}.
The parameter relations \eqref{eq:material_parameter_relation} yields
\begin{align*}
    0 
        &< \frac{9}{2} C \left(l_0 - \bar{l}_0\right) - k_3^2
         = \frac{81}{128}\left( \left( 8\hat{a} + 486\varpi \right)\left( 2\kappa - \bar{\kappa} \right) 
                                - 81\kappa\bar{\kappa}\left( \meanc_{0}- \hat{\meanc}_0 \right)^2 \right)\formPeriod
\end{align*}
Since the leading term is strictly positive, this inequality is satisfied for all special cases in which the second term vanishes,
\eg\ neglecting the Gaussian curvature ($ \bar{\kappa}=0 $) or $ \beta $-independent spontaneous curvature ($ \meanc_{0} = \hat{\meanc}_0 $). 
In the general case, it is most natural to further restrict the double-well parameters $ \hat{a},\varpi > 0 $ rather than the curvature parameters 
$ \kappa\ge\bar{\kappa}\ge 0 $, and $ \meanc_{0}, \hat{\meanc}_0\in\R $.
This leads to
\begin{align}\label{eq:DW_parameter_regime}
    6\varpi + \frac{8}{81}\hat{a}
        &> \frac{\kappa\bar{\kappa}}{2\kappa - \bar{\kappa}}\left( \meanc_{0}- \hat{\meanc}_0 \right)^2 > 0\formPeriod
\end{align}
It should be noted that this is not a substantive restriction, as both double-well parameters can be rescaled simultaneously, 
which only changes the double-well potential \eqref{eq:DW} by an overall prefactor,
\ie\ it is $ f_{\gamma\varpi,\gamma\hat{a}} = \gamma f_{\varpi,\hat{a}}  $ valid for all rescaling factors $ \gamma > 1 $.

\section{Variations}\label{sec:variations}

\subsection{Introduction}

Molecular force fields $ \omega_{\bullet} \in\tangentScal $, fluid force fields $ \Fb_{\bullet} \in\tangentR $, 
and fluid stress fields $ \Sigmab_{\bullet}\in\tangentR\otimes\tangentS < \tangentR[^2] $ are mostly defined in a usual way by
\begin{align*}
    \innerH{\tangentScal}{\omega_{\bullet}, \psi}
        &:= -\frac{2}{3}\innerH{\tangentScal}{\deltafrac{\potenergy_{\bullet}}{\beta}, \psi} \formComma
    & \innerH{\tangentR}{\Fb_{\bullet}, \Wb}
        &:= -\innerH{\tangentR}{\deltafrac{\potenergy_{\bullet}}{\para}, \Wb}\formComma \\
    \innerH{\tangentScal}{\omega_{\bullet}, \psi}
        &:= -\frac{2}{3}\innerH{\tangentScal}{\deltafrac{\fluxpotential_{\bullet}}{\dot{\beta}}, \psi} \formComma
    & \innerH{\tangentR}{\Fb_{\bullet}, \Wb}
        &:= -\innerH{\tangentR}{\deltafrac{\fluxpotential_{\bullet}}{\Vb}, \Wb}\formComma \\
   &&  \innerH{\tangentR[^2]}{\Sigmab_{\bullet}, \nablaC\Wb}
         &:=  \innerH{\tangentR}{\Fb_{\bullet}, \Wb}    
\end{align*}
for all virtual displacements $ \psi\in\tangentScal $, and $ \Wb\in\tangentR $, see \citet{NitschkeVoigt_AiDE_2025}, \ie\ it holds
\begin{align}\label{eq:force_bullet}
    \Fb_{\bullet}
        &= \fb_{\bullet} + \fnor[\bullet]\normal
         = \DivC \Sigmab_{\bullet}
         = \DivC \left(\sigmab_{\bullet} + \normal\otimes\etab_{\bullet} \right)
         = \div\sigmab_{\bullet} - \shop\etab_{\bullet} + \left( \div\etab_{\bullet} + \shop\dbdot\sigmab_{\bullet} \right)\normal\formComma
\end{align}
where $ \sigmab_{\bullet}\in\tangentS[^2] $ are the tangential stress fields,
$ \etab_{\bullet}\in\tangentS  $ the (tangential-)normal stress fields,
$ \fb_{\bullet}\in\tangentS $ the tangential force fields, 
$ f_{\bullet}^{\bot}\in\tangentScal $ the normal force fields,
and $ \bullet $ is a wildcard for the respective name indices, implicitly given by either a potential energy $ \potenergy_{\bullet} $ or a flux potential $ \fluxpotential_{\bullet} $.
The factor $ \frac{2}{3} $ in the definition of scalar order forces $ \omega_{\bullet} $ is chosen for consistency with \citet{NitschkeVoigt_AiDE_2025, NitschkeVoigt_PotRSAMPaES_2025}
and does not affect the scalar order equation $ 0 = \sum_{\bullet} \omega_{\bullet} $.
Essentially, we merely evaluate this equation covariantly rather than contravariantly with respect to the lipid basis 
$ \left( \normal\otimes\normal - \frac{1}{2}\IdS \right) $ used in \eqref{eq:QAnsatz}.
The fluid equations are, according to the Lagrange--D'Alembert principle \citep{NitschkeVoigt_AiDE_2025}, 
given by $ \rho\Ab = \sum_{\bullet}\Fb_{\bullet}  $, 
respectively $ \rho\ab = \sum_{\bullet}\fb_{\bullet} $ in tangential and $ \rho\anor = \sum_{\bullet}\fnor[\bullet] $ in normal direction,
where $ \Ab = \ab + \anor\normal = \Dmat\Vb $ is the material acceleration, see \eqref{eq:accelaration_split} or \cref{tab:forces_conforming}.
Mass conservation is reduced to inextensibility ($ \DivC\Vb = 0 $) in the usual manner \citep{ArroyoDeSimone_PRE_2009, NitschkeVoigt_AiDE_2025} by introducing a Lagrange multiplier, which plays the role of a pressure.
For the sake of simplicity, we neglect all gauges of surface independence \citep{NitschkeSadikVoigt_IJoAM_2023} (\oeda), \ie\ $ \eth_{\Wb}\beta = 0 $ for all surface deformation directions $ \Wb\in\tangentR $.
The governing equations derived from the Lagrange-D'Alembert principle are invariant under any consistent choice of a priori independences between order parameter $ \beta $ and surface $ \surf $, \cf\ \citet{NitschkeVoigt_AiDE_2025}.

\subsection{Surface Landau--Helfrich energy}

In this section we variate the surface Landau--Helfrich energy \eqref{eq:LH}
\begin{align*}
    \energyLH
       	&= \int_{\surf}
                    u_1 
                    + \frac{u_{\kappa}}{2}\left( \meanc - u_{0} \right)^2 
					-u_{\bar{\kappa}}\gaussc
     				+ u_{\text{dw}}
       			\dS\\
\text{with }\quad
    u_0 
        &= f_{\meanc_{0},\hat{\meanc}_{0}}(\beta) = \hat{\meanc}_{0} + \frac{3}{2} ( \meanc_{0} - \hat{\meanc}_{0} )\beta \formComma
    &u_1 
        &= \frac{3}{4} L \normsq{\tangentS}{\nabla\beta} \formComma\\
    u_{\kappa}
        &= f_{\kappa}(\beta) = \frac{9}{4}\kappa\beta^2 \formComma
    &u_{\bar{\kappa}}
        &= f_{\bar{\kappa}}(\beta) = \frac{9}{4}\bar{\kappa}\beta^2 \formComma\\
    u_{\text{dw}}
        &=f_{\varpi,\hat{a}}(\beta) 
                = \frac{9}{4}\beta^2\left( \frac{\hat{a}}{9}\left( \frac{2}{3}-\beta \right)^2 
                    - \frac{3\varpi}{2}\beta\left( 4-\frac{9}{2}\beta \right) \right) \formPeriod
\end{align*}
Except for the $ u_{\bar{\kappa}} $ term, we can fully make use of \citet{BachiniKrauseEtAl_JoFM_2023}. 
The difference is merely semantic: instead of a density function, we have a scalar order parameter.

For $ \potenergy_1 := \int_{\surf} u_1 \dS $,  \citet{BachiniKrauseEtAl_JoFM_2023} shows that we obtain
\begin{align}\label{eq:LH1_force_stress}
\begin{aligned}
    \omega_1 
        &= L \Delta\beta \formComma
    &\sigmab_1
         &= -\frac{3}{2} L \left( \nabla\beta\otimes\nabla\beta - \frac{\normsq{\tangentS}{\nabla\beta}}{2}\IdS \right) \formComma
    &\etab_1
        &= \nullb  \formComma \\
    \fb_1
        &= -\frac{3}{2}\omega_1 \nabla\beta\formComma
    &\fnor[1]
        &= -\frac{3}{2} L \left( \nabla\beta\shop\nabla\beta - \frac{\meanc}{2}\normsq{}{\nabla\beta} \right)\formPeriod
\end{aligned}
\end{align}

For $  \potenergy_0 := \int_{\surf} \frac{u_{\kappa}}{2}\left( \meanc - u_{0} \right)^2  \dS  $,  \citet{BachiniKrauseEtAl_JoFM_2023} also yields
\begin{align}\label{eq:LH0_force_stress}
\begin{aligned}
    \frac{3}{2}\omega_0
        &= - \frac{\partial_\beta u_{\kappa}}{2}\left( \meanc - u_{0} \right)^2
           + u_{\kappa} (\partial_\beta u_{0})\left( \meanc - u_{0} \right) \formComma
    &\sigmab_0
        &= - u_{\kappa}\left( \meanc - u_{0} \right)\left( \shop - \frac{\meanc - u_{0}}{2}\IdS \right) \formComma\\
    \fb_0
        &= -\frac{3}{2}\omega_0 \nabla\beta\formComma
    &\etab_0
        &= -\nabla\left( u_{\kappa} \left( \meanc - u_{0} \right) \right) \formComma\\
    \fnor[0]
        &= -\Delta\left( u_{\kappa} \left( \meanc - u_{0} \right) \right)
           -\frac{u_{\kappa}}{2} \left( \meanc - u_{0} \right)\left( \meanc\left( \meanc + u_{0} \right) - 4\gaussc \right)\formPeriod
\end{aligned}
\end{align}
Expanding the functions $ u_{\kappa}$ and $ u_{0} $, and evaluate all derivatives, leads to
\begin{align}\label{eq:LH0_force_stress_expanded}
\begin{aligned}
    \omega_0
        &= -\frac{3}{2}\kappa \beta\left( \meanc - \hat{\meanc}_{0} - \frac{3}{2} ( \meanc_{0} - \hat{\meanc}_{0} )\beta \right)
                              \left( \meanc - \hat{\meanc}_{0} - 3 ( \meanc_{0} - \hat{\meanc}_{0} )\beta \right) \formComma\\
     \sigma_0
        &= -\frac{9}{8}\kappa\beta^2 \left( \meanc - \hat{\meanc}_{0} - \frac{3}{2} ( \meanc_{0} - \hat{\meanc}_{0} )\beta \right)
                              \left( 2\shop - \left( \meanc - \hat{\meanc}_{0} - \frac{3}{2} ( \meanc_{0} - \hat{\meanc}_{0} )\beta \right)\IdS \right)\formComma\\
    \etab_0
        &= -\frac{9}{4}\kappa\beta \left( \beta\nabla\meanc
                                        +2\left( \meanc - \hat{\meanc}_{0} - \frac{9}{4} ( \meanc_{0} - \hat{\meanc}_{0} )\beta \right)\nabla\beta \right)\formComma\\
    \fnor[0]
        &= --\frac{9}{4}\kappa\Bigg( \beta^2\Delta\meanc 
                                    + 2\beta\left( \meanc - \hat{\meanc}_{0} - \frac{9}{4} ( \meanc_{0} - \hat{\meanc}_{0} )\beta \right)\Delta\beta\\
                      &\hspace{4em} + 2\inner{\tangentS}{2\beta\nabla\meanc 
                                                         + \left( \meanc - \hat{\meanc}_{0} - \frac{9}{2} ( \meanc_{0} - \hat{\meanc}_{0} )\beta \right)\nabla\beta
                                                                , \nabla\beta}\\
                      &\hspace{4em} + \frac{\beta^2}{2} \left( \left( \meanc^2 
                                                                       -  \left( \hat{\meanc}_{0} + \frac{3}{2} ( \meanc_{0} - \hat{\meanc}_{0} )\beta \right)^2\right)\meanc
                                                               -4\left( \meanc - \hat{\meanc}_{0} - \frac{3}{2} ( \meanc_{0} - \hat{\meanc}_{0} )\beta \right)\gaussc 
                                                         \right)
                              \Bigg) \formPeriod
\end{aligned}
\end{align}

Unfortunately, the energy part $  \potenergy_{\bar{\kappa}} := -\int_{\surf} u_{\bar{\kappa}}\gaussc  \dS  $ is not treated in \citet{BachiniKrauseEtAl_JoFM_2023} in any way.
Therefore, we derive the variations of this energy in the following.
The deformation derivative $ \eth_{\Wb}\gaussc $ of the Gaussian curvature $ \gaussc $ in deformation direction $ \Wb\in\tangentR $ is given in \citet{SischkaNitschkeEtAl_FD_2025}:
\begin{align*}
    \eth_{\Wb}\gaussc
        &= \inner{\tangentS[^2]}{\meanc\IdS - \shop,  \nabla(\normal\nablaC\Wb)}
           -\gaussc\DivC\Wb \formPeriod
\end{align*}
This and $ \eth_{\Wb}\beta \equiv 0 $, \oeda, leads to the spatial variation
\begin{align*}
    \innerH{\tangentR}{\deltafrac{\potenergy_{\bar{\kappa}}}{\para}, \Wb}
        &= -\int_{\surf} u_{\bar{\kappa}}\left( \eth_{\Wb}\gaussc + \gaussc\DivC\Wb \right) \dS
         = -\innerH{\tangentS[^2]}{ u_{\bar{\kappa}} \left( \meanc\IdS - \shop \right) , \nabla(\normal\nablaC\Wb) }\\
        &= \innerH{\tangentR[^2]}{ \normal \otimes \div\left( u_{\bar{\kappa}} \left( \meanc\IdS - \shop \right) \right), \nablaC\Wb } \formComma
\end{align*}
\ie, with $ \div(\meanc\IdS - \shop) = \nullb $, we obtain
\begin{align}\label{eq:LHbarkappa_stress}
    \sigmab_{\bar{\kappa}} 
        &= \nullb \formComma
    &\etab_{\bar{\kappa}}
        &= \frac{9}{2}\bar{\kappa} \beta \left( \meanc\IdS - \shop \right) \nabla\beta 
\end{align}
for the associated stress.
Forces are given by
\begin{align}\label{eq:LHbarkappa_force}
\begin{aligned}
    \omega_{\bar{\kappa}}
        &= \frac{2}{3}(\partial_\beta u_{\bar{\kappa}}) \gaussc
         = 3\bar{\kappa} \gaussc \beta \formComma\\
    \fb_{\bar{\kappa}}
        &= -\shop\etab_{\bar{\kappa}}
         = -\frac{9}{2}\bar{\kappa} \gaussc \beta \nabla\beta
         = -\frac{3}{2}\omega_{\bar{\kappa}} \nabla\beta \formComma \\
    \fnor[\bar{\kappa}]
        &= \div\etab_{\bar{\kappa}}
         = -\frac{9}{2}\bar{\kappa} \left( \beta \shop\dbdot\nabla^2\beta - \meanc\beta\Delta\beta 
                                            + (\nabla\beta)\shop\nabla\beta - \meanc \normsq{\tangentS}{\nabla\beta}\right) \formComma
\end{aligned}
\end{align}
where we used that $ \shop^2 = \meanc\shop - \gaussc\IdS $ is valid.

For the double-well potential $  \potenergy_{\text{dw}} := \int_{\surf} u_{\text{dw}}  \dS  $, 
we can once more employ \citet{BachiniKrauseEtAl_JoFM_2023} and obtain
\begin{align}\label{eq:LHdw_stress_force}
\begin{aligned}
    u_{\text{dw}}
            &= \frac{9}{4}\beta^2\left( \frac{\hat{a}}{9}\left( \frac{2}{3}-\beta \right)^2 
                        - \frac{3\varpi}{2}\beta\left( 4-\frac{9}{2}\beta \right) \right) \formComma
    &\sigmab_{\text{dw}}
        &= u_{\text{dw}}\IdS \formComma\\
    \omega_{\text{dw}}
        &= -\frac{2}{3}\partial_\beta u_{\text{dw}}
         = -\frac{1}{6}\beta\left( \frac{2}{3} - \beta \right)\left( 4 \hat{a}\left( \frac{1}{3}-\beta \right)-243\varpi\beta \right) \formComma
    &\etab_{\text{dw}}
        &= \nullb\formComma\\
    \fb_{\text{dw}}
        &= \nabla u_{\text{dw}}
         = -\frac{3}{2}\omega_{\text{dw}}\nabla\beta\formComma
    &\fnor[\text{dw}]
        &= \meanc u_{\text{dw}}\formPeriod
\end{aligned}
\end{align}
It should be noted that the fluid force $ \Fb_{\text{dw}}=\fb_{\text{dw}} +  \fnor[\text{dw}]\normal $ is not an applied force in inextensible materials 
and could be neglected,
since $ \Fb_{\text{dw}} = \GradC u_{\text{dw}} $ is a pure pressure force.

\subsection{Flux potentials}

As demonstrated in \cref{sec:monolayer}, the lipid ansatz \eqref{eq:QAnsatz} can be used without considering additional lipid constraint forces. 
Consequently, the flux forces and stresses of the surface conforming model in \cref{sec:full_surface_conforming_models} may also be evaluated with $ \qb=\nullb $.
However, it is useful to explicitly write down the flux potentials resulting from the ansatz \eqref{eq:QAnsatz} for a better understanding, and to employ the corresponding variations as an additional means of verification, since ``Variation$ \circ $Restriction'' is commutative in the absence of constraints forces.
Nevertheless, we will keep the following calculations brief.

By the lipid ansatz \eqref{eq:QAnsatz} the anisotropic fluid metric yields
\begin{align*}
    \Ib_{\xi}[\beta] 
        &:= \Ib_{\xi}[\Qb] 
         = \Id - \xi \Qb
         = \left(1+\frac{\xi}{2}\beta\right)\IdS + \left( 1-\xi\beta \right)\normal\otimes\normal \formComma
\end{align*}
where $ \xi\in\R $ is the anisotropy coefficient.
Its temporal distortion is given by the lower-convected rate \citep{NitschkeVoigt_JoGaP_2023}
\begin{align} \label{eq:DlowIbxi}
    \Dlow\Ib_{\xi}[\beta]
        &= \left(1+\frac{\xi}{2}\beta\right)\IdS\left( \nablaC\Vb + \nablaC^T\Vb \right)\IdS
          -\xi\dot{\beta} \left( \normal\otimes\normal - \frac{1}{2}\IdS \right)
\end{align}
and determine the heat released by material deformations modulo rigid body motions.
The associated viscous flux potential $ \energyNV = \frac{\coeffIF}{4} \normHsq{\tangentR[^2]}{\Dlow\Ib_{\xi}[\beta]} $ reads
\begin{align}\label{eq:visc_flux_pot}
    \energyNV
        &= \frac{\coeffIF}{8} \Bigg( 2 \normHsq{\tangentS[^2]}{\left(1+\frac{\xi}{2}\beta\right)\IdS\left( \nablaC\Vb + \nablaC^T\Vb \right)\IdS} \\
                     &\hspace{3em}+ 3 \xi^2 \normHsq{\tangentScal}{\dot{\beta}}
                                    + 4 \xi \innerH{\tangentScal}{\left(1+\frac{\xi}{2}\beta\right)\dot{\beta}, \DivC\Vb }\Bigg)\formComma \notag
\end{align}
where $ \coeffIF \ge 0 $ is the isotropic viscosity.
Variation \wrt\ process variables $ \dot{\beta} $ and $ \Vb=\vb + \vnor\normal $ yields
\begin{align}\label{eq:NV_force_stress}
\begin{aligned}
    \omegaNV
        &= - \xi\coeffIF\left( \frac{\xi}{2} \dot{\beta} + \frac{1}{3}\left(1+\frac{\xi}{2}\beta\right)\left( \div\vb - \meanc\vnor \right) \right) \formComma\\
    \sigmabNV
        &= \coeffIF\left( \left(1+\frac{\xi}{2}\beta\right)^2\left( \nabla\vb + \nabla^T\vb - 2\vnor\shop \right) 
                                                                     + \frac{\xi}{2}\dot{\beta}\left(1+\frac{\xi}{2}\beta\right)\IdS\right)\formComma
    &\etab_{\NV}
            &= \nullb \formComma             
\end{aligned}
\end{align}
which equals the viscosity contributions in the surface-conforming model in \cref{sec:full_surface_conforming_models}, \resp\ \citet{NitschkeVoigt_AiDE_2025}, for $ \qb=\nullb $. 
It should be noted that, in inextensible materials, the trailing summands can be neglected since they either vanish or lead to a 
pressure gradient.
Note that for comparisons with \cref{tab:forces_conforming} it holds
$ \omegaNV = \xi^2(\tilde{\omega}^2_{\NV} - \frac{\coeffIF}{2}\dot{\beta}) $
and $ \sigmabNV = \sigmabNV^0 + \xi\sigmabNV^1 +  \xi^2\sigmabNV^2 $.
Although we would not recommend proceeding in this way, for completeness the tangential and normal fluid forces for isotropic viscous fluids ($\xi=0$) are
\begin{align*}
    \fb_{\text{iso}} 
    	&:=\coeffIF \div\left( \nabla\vb + \nabla^T\vb - 2\vnor\shop \right)
         = \coeffIF \left(\Delta\vb + \gaussc\vb - 2\left( \shop - \frac{\meanc}{2}\IdS \right)\nabla\vnor 
         	- \vnor\nabla\meanc + \nabla\left( \div\vb - \vnor\meanc \right)\right) \formComma\\
    \fnor[\text{iso}]
    	&:= \coeffIF\shop\dbdot\left( \nabla\vb + \nabla^T\vb - 2\vnor\shop \right) 
    	  = 2\coeffIF\left( \shop\dbdot\nabla\vb - \vnor\left( \meanc^2 - \gaussc \right) \right)\formComma
\end{align*}
see \citet{ArroyoDeSimone_PRE_2009,ReutherNitschkeEtAl_JoFM_2020},
and  yield the more general forces
\begin{align}\label{eq:NV_fluid_force}
\begin{aligned}
    \fb_{\NV}
    	&=  \left(1+\frac{\xi}{2}\beta\right)^2 \fb_{\text{iso}} 
    		+\coeffIF \left(\xi \left(1+\frac{\xi}{2}\beta\right)\left( \nabla\vb + \nabla^T\vb - 2\vnor\shop \right)\nabla\beta 
    						+\frac{\xi^2}{4}\dot{\beta}\nabla\beta + \frac{\xi}{2}\left(1+\frac{\xi}{2}\beta\right)\nabla\dot{\beta} \right)\formComma\\
  	\fnor[\NV]
  		&= \left(1+\frac{\xi}{2}\beta\right)^2\fnor[\text{iso}]
  			+\frac{\xi\coeffIF}{2}\dot{\beta}\left(1+\frac{\xi}{2}\beta\right)\meanc \formPeriod
\end{aligned}					
\end{align}

Next, we address the immobility flux potential $ \energyIM = \frac{M}{2}\normHsq{\tangentR[^2]}{\Dt[\Phi]\Qb} $ with $ \Phi\in\{\mfrak,\jau\} $, where
\begin{align*}
    \Dmat\Qb
        &= \dot{\beta}\left( \normal\otimes\normal - \frac{1}{2}\IdS \right)
            -\frac{3}{2}\beta\left( \normal\nablaC\Vb \otimes \normal + \normal \otimes \normal\nablaC\Vb  \right) \formComma
    &\Djau\Qb
        &= \dot{\beta}\left( \normal\otimes\normal - \frac{1}{2}\IdS \right)
\end{align*}
holds by \citet{NitschkeVoigt_JoGaP_2023} and lipid ansatz \eqref{eq:QAnsatz}.
This results in
\begin{align}\label{eq:immobility_flux_pot}
    \energyIM[\mfrak]
        &= \frac{3 M}{4}\left( \normHsq{\tangentScal}{\dot{\beta}} + 3 \normHsq{\tangentS}{\beta\normal\nablaC\Vb} \right)\formComma
    &\energyIM[\jau]
        &= \frac{3 M}{4} \normHsq{\tangentScal}{\dot{\beta}} \formPeriod
\end{align}
Variation \wrt\ process variables $ \dot{\beta} $ and $ \Vb=\vb + \vnor\normal $ yields
\begin{align}\label{eq:IM_force_stress}
\begin{aligned}
    \omegaIM[\mfrak] 
        &= \omegaIM[\jau]
         = -M \dot{\beta} \formComma\\
    \sigmabIM[\mfrak]
        &= \sigmabIM[\jau]
         = \nullb \formComma
    &\etab_{\IM}^{\mfrak}
        &= \frac{9 M}{2} \beta^2\left( \nabla\vnor + \shop\vb \right) \formComma
    &\etab_{\IM}^{\jau}
        &= \nullb \formPeriod
\end{aligned}
\end{align}
This also agrees with the surface-conforming model in \cref{sec:full_surface_conforming_models}, \resp\ \citet{NitschkeVoigt_AiDE_2025}, for $ \qb=\nullb $. 
It should be noted that the surface-conforming model does not impose any normal immobility stresses $ \etab_{\IM}^{\Phi} $ directly.
Instead, it comprises constraint forces to ensure surface-conformity, \eg\ in terms of $ \zetabIM\in\tangentS $.
However, the lipid ansatz \eqref{eq:QAnsatz} already stipulates surface-conformity a priori.
Both procedures are equivalent eventually, and for $ \qb = \nullb $ it holds $ \etab_{\IM}^{\Phi} = 3\beta \zetabIM$ and thus gives the same immobility force in the fluid equation.
For the sake of completeness, the resulting fluid forces in tangential and normal directions are
\begin{align}\label{eq:IM_fluid_force}
\begin{aligned}
    \fb_{\IM}^{\mfrak}
    	&= \frac{9 M}{2} \beta^2\left( \gaussc\vb - \shop\left( \nabla\vnor + \meanc\vb \right) \right)\formComma
    &\fb_{\IM}^{\jau}
    		&= \nullb\formComma\\
    f^{\mfrak,\bot}_{\IM}
    	&= \frac{9 M}{2} \beta \left( \left( \Delta\vnor + \shop\dbdot\nabla\vb + \nabla_{\vb}\meanc \right)\beta 
    									+ 2\left( \nabla\vnor + \vb\shop \right)\nabla\beta\right) \formComma
    &f^{\jau,\bot}_{\IM}
        	&= 0 \formPeriod
\end{aligned}
\end{align}

The activity flux potential $ \energyAC = \frac{\alpha}{2}\int_{\surf}(\Tr\circ\Dlow\Ib_{\xi}[\Qb])\dS $, which stated the first moment of the temporal anisotropic metric distortion, is not relevant for lipids on an inextensible medium.
This can be seen by \eqref{eq:DlowIbxi}, which yields
\begin{align*}
    \energyAC = \alpha \int_{\surf} \left( 1 + \frac{\xi}{2}\beta \right) \DivC\Vb \dS \formPeriod
\end{align*}
Variations results in
\begin{align}\label{eq:AC_stress_force}
\begin{aligned}
    \omega_{\AC} &= 0 \formComma
    &\sigmabAC	&= \alpha \left( 1 + \frac{\xi}{2}\beta \right) \IdS \formComma
    & \etab_{\AC} &= \nullb \formComma\\
    &
    &\fb_{\AC} &= \alpha\nabla\left( 1 + \frac{\xi}{2}\beta \right) = \frac{\alpha\xi}{2}\nabla\beta \formComma
    &\fnor[\AC] &= \alpha \left( 1 + \frac{\xi}{2}\beta \right)\meanc\formPeriod
\end{aligned}
\end{align}
That means we only get an active pressure term, which has not any influence of the solution for an inextensible fluid and thus can be neglected. 
Although it is irrelevant for the present paper, it should be noted that $ \sigmabAC $ gives rise to geometric activity \citep{NitschkeVoigt_PotRSAMPaES_2025},
or active tension \citep{MietkeJemseenaEtAl_PRL_2019,WittwerAland_JoCPX_2023},
\resp\ contributes to the membrane tension \citep{Al-IzziAlexander_PRR_2023},
depending on the scalar order in an extensible medium.

\section{Energy rate}\label{sec:energy_rate}

In this section we discuss the energy energy rate $ \ddt\energy_{\TOT} $ of the total energy
\begin{align*}
    \energy_{\TOT}
        &:= \int_{\surf} \frac{\rho}{2} \normsq{\tangentR}{\Vb} \dS
            + \energyLH
\end{align*}
consisting of the kinetic energy and the potential energy, the latter being given solely by the LH energy \eqref{eq:LH}.
Conceptually, one may either appeal to the intermediate result in \citet{NitschkeVoigt_AiDE_2025} for the energy rate under the ansatz \eqref{eq:QAnsatz}, 
or one may re-derive it following the same procedure as in \citet{NitschkeVoigt_AiDE_2025} within our framework. 
Either way, the Lagrange--d'Alembert principle yields the physical power balance
\begin{align*}
    \ddt\energy_{\TOT}
        &= \power_{\TOT}
         = \power_{\NV} + \power_{\IM}^{\Phi} + \power_{\AC}
\end{align*}
for solutions of the governing equations \eqref{eq:full_R3},
where the individual power inputs $ \power_{\bullet} $, due to anisotropic viscosity, immobility, and activity fluxes 
are given by the product of 
generalized forces $ \left(\Fb_{\bullet}, \Hb_{\bullet}\right) \in \tangentR\times\tangentQR $, 
\resp\ $ \left( \fb_{\bullet}, \fnor[\bullet], \omega_{\bullet} \right)\in\tangentS\times\tangentScal\times\tangentScal $, 
and generalized velocities $ \left( \Vb, \Dmat\Qb \right) \in \tangentR\times\tangentQR $,
\resp\ $ \left( \vb, \vnor, \dot{\beta} \right)\in\tangentS\times\tangentScal\times\tangentScal $:
\begin{align*}
    \power_{\bullet}
        &= \innerH{\tangentR}{\Fb_{\bullet}, \Vb}
            + \innerH{\tangentQR}{\Hb_{\bullet}, \Dmat\Qb}
         = \innerH{\tangentS}{\fb_{\bullet}, \vb}
            + \innerH{\tangentScal}{\fnor[\bullet], \vnor}
            + \frac{3}{2}\innerH{\tangentScal}{\omega_{\bullet}, \dot{\beta}} \\
        &= -\left( \innerH{\tangentS[^2]}{\sigmab_{\bullet}, \nabla\vb - \vnor\shop} 
                   +\innerH{\tangentS}{\etab_{\bullet}, \nabla\vnor + \shop\vb}
                   - \frac{3}{2}\innerH{\tangentScal}{\omega_{\bullet}, \dot{\beta}}\right) 
\end{align*}
with fluid stress fields $ \left( \sigmab_{\bullet}, \etab_{\bullet} \right)\in\tangentS[^2]\times\tangentS $ \eqref{eq:force_bullet}. 
Inextensibility \eqref{eq:full_R3_konti}, and stress/force \eqref{eq:NV_force_stress} result in $ \power_{\NV} = -2\energyNV $ \eqref{eq:visc_flux_pot};
stress/force \eqref{eq:IM_force_stress} yield $ \power_{\IM}^{\Phi} = -2 \energyIM[\Phi] $ \eqref{eq:immobility_flux_pot} for $ \Phi\in\{\mfrak, \jau\} $;
the activity power/flux potential $ \power_{\AC}= \energyAC = 0 $ vanish as expected by \eqref{eq:AC_stress_force}.
Eventually, we summarize
\begin{align*}
    \ddt\energy_{\TOT} 
        &= - \frac{1}{4}\Bigg(  2\coeffIF \normHsq{\tangentS[^2]}{\left(1+\frac{\xi}{2}\beta\right)\IdS\left( \nablaC\Vb + \nablaC^T\Vb \right)\IdS} \\ 
                      &\hspace{3em}+ 3\left( 2M + \coeffIF \xi^2\right) \normHsq{\tangentScal}{\dot{\beta}} 
                                             + 18M \delta_{\mfrak}^{\phi} \normHsq{\tangentS}{\beta\normal\nablaC\Vb}\Bigg) \notag\\
        &= -2 \left( \energyNV + \energyIM[\Phi] \right) \label{eq:energyrate}
        \le 0
\end{align*}
for solutions of the LH model \eqref{eq:full_R3}.
Consequently, the evolution of both the surface $ \surf $ and the scalar order parameter $ \beta $ is of a purely dissipative nature. 
This dissipation is irreversible and hence directly proportional to the entropy production.

\section{Surface conforming Beris--Edwards model}\label{sec:full_surface_conforming_models}

In the following we merely recall the surface-conforming active Beris--Edwards model from \citet{NitschkeVoigt_PotRSAMPaES_2025}. 
The full derivation and all mathematical background that is not reiterated here are provided in \citet{NitschkeVoigt_AiDE_2025, NitschkeVoigt_PotRSAMPaES_2025}.\\

\begin{model}
    Find the tangential and normal material velocity fields $ \vb\in\tangentS $ and $ \vnor\in\tangentS[^0] $, tangential Q-tensor field $ \qb\in\tangentQS $, normal eigenvalue field $ \beta\in\tangentS[^0] $, pressure field $ p\in\tangentS[^0] $ and Lagrange parameter fields $ \lambdab_{\gamma}\in\Vcal_{\gamma}\le\tangentR[^n] $ for all $ \gamma\in\Cset_{\SC} $ \soth
    \begin{subequations}\label{eq:model_conforming}
       \begin{gather}
       \begin{align}
            \rho\ab 
                &= \nabla\tp   + \div\left( \sigmabEL + \sigmabIM  + \sigmabNV^0 + \xi\sigmabNV^1 +  \xi^2\sigmabNV^2 +\sigmabNA\right) \notag\\
                &\quad + \left(2\shop\qb-3\beta\shop\right)\left( \zetabEL + \zetabIM \right) \label{eq:model_conforming_tangentialeq}
                       + \sum_{\gamma\in\Cset_{\SC}}\fb_{\gamma}\formComma\\
            \rho\anor 
                &= \tp\meanc + \fnorBE  + \shop\dbdot\left( \sigmabEL + \sigmabIM  + \sigmabNV^0 + \xi\sigmabNV^1 + \xi^2\sigmabNV^2 + \sigmabNA \right) \notag\\
                &\quad - \div\left( \left(2\qb-3\beta\IdS\right) \left( \zetabEL + \zetabIM \right) \right)
                       + \sum_{\gamma\in\Cset_{\SC}} \fnor[\gamma] \formComma \label{eq:model_conforming_normalseq}
       \end{align}\\
       \begin{align}
            \left( M + \frac{\coeffIF\xi^2}{2} \right)\timeD\qb \label{eq:model_conforming_qtensor}
                &= \hbEL + \hbTH + \xi\hbNV^1 + \xi^2\widetilde{\hb}^{2,\Phi}_{\NV} + \sum_{\gamma\in\Cset_{\SC}} \hb_{\gamma} \formComma \\
            \left( M + \frac{\coeffIF\xi^2}{2} \right) \dot{\beta} \label{eq:model_conforming_betaeq}
                &= \omegaEL + \omegaTH  + \xi^2\widetilde{\omega}^{2}_{\NV} + \sum_{\gamma\in\Cset_{\SC}} \omega_{\gamma}\formComma
       \end{align}\\
       \div\vb = \vnor\meanc \formComma\\
       \forall\gamma\in\Cset_{\SC}:\quad  \nullb = \Cb_{\gamma} \label{eq:model_conforming_constraineqs}
       \end{gather}
    \end{subequations}
    holds for $ \dot{\rho}=0 $ and given initial conditions for $ \vb $, $ \vnor $, $ \qb $, $ \beta $ and mass density $ \rho\in\tangentS[^0] $. \\
\end{model}

Quantities are unpacked in \cref{tab:forces_conforming}. 
The choice of $ \Phi $ for $  \sigmabNV^1  $ and $  \sigmabNV^2  $ is optional. Both given representations state equal tensor fields.
$ \Cset_{\SC} $ is a set of constraints suitable for surface conforming Q-tensor fields, which might yield additional constraint forces.
In this paper we only consider the ``No flat Degeneracy'' ($ \ND $) constraint, \ie\ $ \Cset_{\SC} = \{\ND\} $ holds and
$ \fb_{\ND}, \fnor[\ND], \hb_{\ND}, \omega_{\ND}, \Cb_{\ND} $ are the associated generalized constraint forces derived in \cref{sec:NoFlatDegeneracy}.
Since this is an inextensible model we summarize obvious pressure terms into the generalized pressure $ \tp\in\tangentScal $ for simplicity.
\begin{table}
    \centering
    \scalebox{0.75}{
    \begin{tabular}{lcr}
        Identifier
            & Expression
                & Naming \\
    \hline
        $ \ab $
            & $ (\partial_t v^i)\partial_i\para_{\!\ofrak} + \nabla_{\vb-\vb_{\!\ofrak}}\vb + \nabla_{\vb}\vb_{\!\ofrak} - \vnor\left( \nabla\vnor + 2\shop\vb \right) $
                & tan.~acceleration  \\
    \hline
        $ \anor $
            & $ \partial_t\vnor + \nabla_{2\vb-\vb_{\!\ofrak}}\vnor + \shop(\vb,\vb) $ 
                & nor.~acceleration\\
    \hline
        $ \timeD[\jau]\qb $
            & $ \timeJ\qb = \dot{\qb} - \Ab[\Vb]\qb + \qb\Ab[\Vb] $
                    & Jaumann time derivative\\
    \hline
        $ \timeD[\mfrak] $
            & $ \dot{\qb} = (\partial_t q^{ij})\partial_i\para_{\!\ofrak}\otimes\partial_j\para_{\!\ofrak} + \nabla_{\vb-\vb_{\!\ofrak}}\qb + \Gb[\Vb_{\!\ofrak}]\qb + \qb\Gb^T[\Vb_{\!\ofrak}] $ 
                    & material time derivative\\
    \hline
        $ \sigmabEL $
             & \splitcelltab{$ -L\big( (\nabla\qb)^{T} \dbdot\nabla\qb + \frac{3}{2}\nabla\beta\otimes\nabla\beta 
                                          - \frac{1}{4}\left( 2\normsq{\tangentS[^3]}{\nabla\qb} + 3\normsq{\tangentS[^2]}{\nabla\beta}  \right)\IdS  $ \\
                                $ - 6\gaussc\beta\qb + \frac{1}{2}\left( 2\meanc\Tr\qb^2 - 12\beta \qb \dbdot \shop  + 9 \meanc\beta^2 \right)\left( \shop - \frac{\meanc}{2}\IdS \right)\big) $}
                    & elastic stress \\
    \hline
        $ \zetabEL $
             & $ L\left( 2(\nabla\qb)\dbdot\shop + \qb\nabla\meanc - 3\shop\nabla\beta - \frac{3}{2}\beta\nabla\meanc \right) $ 
                    & elastic pt.~of surface-conforming constr.\\
    \hline
        $ \hbEL $
             & $ L \left(  \Delta\qb - (\meanc^2-2\gaussc)\qb + 3\beta \meanc \left( \shop - \frac{\meanc}{2}\IdS \right) \right) $
                    & tan.~mol.~elastic force\\
    \hline
        $ \omegaEL $
             & $ L \left( \Delta\beta + 2 \meanc\shop \dbdot \qb  - 3\beta \left( \meanc^2 - 2\gaussc \right)  \right) $ 
                    & nor.~mol.~elastic force\\
    \hline
        $ \hbTH $
             & $ -\left(2a - 2b\beta + 3c\beta^2  + 2c\Tr\qb^2\right)\qb $
                    & tan.~mol.~thermotropic force \\
    \hline
        $ \omegaTH $
             & $ -\left(2a + b\beta + 3c\beta^2  + 2c\Tr\qb^2\right)\beta + \frac{2}{3} b \Tr\qb^2 $ 
                    & nor.~mol.~thermotropic force\\
    \hline
        $ \fnorBE $
             & $ -\kappa\left( \Delta\meanc + \left( \meanc - \meanc_{0} \right)\left( \frac{1}{2}\meanc(\meanc + \meanc_{0}) - 2\gaussc  \right) \right) $
                    & bending force \\
    \hline
        $ \sigmabIM[\jau] $
             & $ M \left( \qb\timeJ\qb - (\timeJ\qb)\qb \right) $
                    & Jaumann immobility stress \\
    \hline
        $ \sigmabIM[\mfrak] $
             & $ \nullb $ 
                    & material immobility stress\\
    \hline
        $ \zetabIM[\jau] $
             & $ \nullb $ 
                    & Jaumann immobility pt.~of surface-conforming constr.\\
    \hline
        $ \zetabIM[\mfrak] $
             & $  -M \left(\nabla\vnor + \vb\shop\right)\left( \qb - \frac{3}{2}\beta\IdS \right) $ 
                    & material immobility pt.~of surface-conforming constr.\\
    \hline
        $ \sigmabNV^0 $
             & $ 2\coeffIF\Sb[\Vb] $
                    & iso.~viscosity stress\\
    \hline
        $ \sigmabNV^1 $ ($\jau$-form)
             & $ -\coeffIF\left(\timeJ\qb - \frac{\dot{\beta}}{2}\IdS + 3\qb\Sb[\Vb] + \Sb[\Vb]\qb - 2\beta\Sb[\Vb] \right) $
                    & 1st ord.~aniso.~viscosity stress\\
    \hline
        $ \sigmabNV^1 $ ($\mfrak$-form)
             & $ -\coeffIF\left( \dot{\qb} - \frac{\dot{\beta}}{2}\IdS + \qb(2\Gb[\Vb]+\Gb^T[\Vb]) + \Gb^T[\Vb]\qb - 2\beta\Sb[\Vb] \right)  $
                    & 1st ord.~aniso.~viscosity stress\\
    \hline
        $ \hbNV^1 $
             & $ \coeffIF\Sb[\Vb] $ 
                    & 1st ord.~mol.~aniso.~viscosity force\\
    \hline
        $ \sigmabNV^2 $ ($\jau$-form)
                & \splitcelltab{$ \coeffIF\big( \qb\timeJ\qb - \frac{1}{2}\left( \beta\timeJ\qb + \dot{\beta}\qb \right)
                        + \frac{1}{4}\beta\dot{\beta}\IdS + \qb\Sb[\Vb]\qb$ \\
                  $-\frac{1}{2}\beta\left( 3\qb\Sb[\Vb] + \Sb[\Vb]\qb \right)  
                    + \frac{1}{2}\left( \Tr\qb^2 + \beta^2 \right)\Sb[\Vb] \big) $} 
                        & 2nd ord.~aniso.~viscosity stress\\
    \hline
        $ \sigmabNV^2 $ ($\mfrak$-form)
                & \splitcelltab{$ \coeffIF\big(  \qb\dot{\qb} - \frac{1}{2}\left( \beta\dot{\qb} + \dot{\beta}\qb \right) 
                          + \frac{1}{4}\beta\dot{\beta}\IdS + \qb\Gb^T[\Vb]\qb$ \\
                  $ -\frac{1}{2}\beta\left( \qb(2\Gb[\Vb]+\Gb^T[\Vb]) + \Gb^T[\Vb]\qb \right)  
                  + \frac{1}{2}(\Tr\qb^2)\Gb[\Vb] + \frac{1}{2}\beta^2\Sb[\Vb]\big) $} 
                        & 2nd ord.~aniso.~viscosity stress\\
    \hline
        $ \widetilde{\hb}^{2,\jau}_{\NV} $
                & $ -\frac{\coeffIF}{2} \left( \qb\Sb[\Vb] + \Sb[\Vb]\qb - (\qb\dbdot\Gb[\Vb])\IdS - \beta\Sb[\Vb]\right) $
                    & pt.~of 2nd ord.~mol.~aniso.~viscosity force\\
    \hline
        $ \widetilde{\hb}^{2,\mfrak}_{\NV} $
                & $ -\frac{\coeffIF}{2} \left( \qb\Gb[\Vb] + \Gb^T[\Vb]\qb -(\qb\dbdot\Gb[\Vb])\IdS - \beta\Sb[\Vb] \right) $
                    & pt.~of 2nd ord.~mol.~aniso.~viscosity force\\
    \hline
        $ \widetilde{\omega}^{2}_{\NV} $
                & $ \frac{\coeffIF}{3} \qb\dbdot\Gb[\Vb] $ 
                    & pt.~of 2nd ord.~mol.~aniso.~viscosity force\\
    \hline
        $ \sigmabNA $
            & $ \cNA\qb $
                & nematic activity stress
    \end{tabular}}
    \caption[Necessary terms for Surface Conforming Beris--Edwards models]
       		{Necessary terms for the Surface Conforming Beris--Edwards models \eqref{eq:model_conforming} for a consistent choice $ \Phi\in\{\jau,\mfrak\} $.
            These representations comprise the tangential deformation gradient $ \Gb[\Vb] = \nabla\vb - \vnor\shop $ of the material velocity $ \Vb=\vb+\vnor\normal $.
            $ \Sb[\Vb] $ and $\Ab[\Vb]$ are its symmetric and skew-symmetric part.
            Time derivatives are determined \wrt\ an observer velocity $ \Vb_{\!\ofrak}= \vb_{\ofrak} + \vnor\normal $.
            For more details see \citet{NitschkeVoigt_AiDE_2025,NitschkeVoigt_PotRSAMPaES_2025}.}
    \label{tab:forces_conforming}
\end{table}

\section{No flat degeneracy constraint}\label{sec:NoFlatDegeneracy}

The ``No flat Degeneracy'' ($ \ND $) constraint may be given by
\begin{align}\label{eq:ND_constraint}
    \nullb 
        &= \qb 
         = (\projQS\circ\projS[^2])\Qb
         = \IdS\Qb\IdS-\frac{\Qb\dbdot\IdS}{2}\IdS = \Cb_{\ND} \formComma
\end{align}
where $ \projS[^2]:\tangentR[^2] \rightarrow \tangentS[^2] $ is the orthogonal tangential projection on 2-tensor fields 
with $ \projS[^2]\Rb = \IdS\Rb\IdS $ for all $ \Rb\in\tangentR[^2] $,
and $ \projQS: \tangentS[^2] \rightarrow \tangentQS $ is the orthogonal flat-degenerated Q-tensor projection on tangential 2-tensor fields
with $ \projQS \rb = \frac{1}{2}\left( \rb + \rb^T - (\Tr\rb)\IdS \right) $.
To implement this into the surface conforming model \eqref{eq:model_conforming}, we introduce the Lagrange functional
\begin{align*}
    \energyND 
        := -\innerH{\tangentQS}{\lambdabND,(\projQS\circ\projS[^2])\Qb}
         = -\innerH{\tangentQR}{\lambdabND,\Qb}
\end{align*}
as a potential energy,
with Lagrange parameter $ \lambdabND\in\tangentQS $ yielding an additional degree of freedom. 
The variation 
\begin{align*}
	\innerH{\tangentQS}{\Cb_{\ND}, \thetab}
    &:=-\innerH{\tangentQS}{\deltafrac{\energyND}{\lambdabND}, \thetab} 
      = 0
\end{align*}
for all virtual displacements $ \thetab\in\tangentQS $, results in the constraint equation \eqref{eq:ND_constraint}.
The molecular constraint force $\HbND\in\tangentQR$ is given by the variation
\begin{align*}
 \innerH{\tangentQR}{\HbND, \Psib}
    	&:= -\innerH{\tangentQR}{\deltafrac{\energyND}{\Qb}, \Psib}
  	  =  \innerH{\tangentQR}{\lambdabND, \Psib}
\end{align*}
for all virtual displacements $ \Psib\in\tangentQR $.
Therefore, the constraint force is flat degenerated and we have to add
\begin{align*}
    \hbND
    	:= (\projQS\circ\projS[^2])\HbND
    	 = \HbND
    	 = \lambdabND \in\tangentQS
\end{align*}
to the right-handed side of the flat degenerated molecular equation \eqref{eq:model_conforming_qtensor}.
Since $\lambdabND$ is fully determined by that equation, we can, in principle, omit it.
However, as is the case with the surface conforming constraint in \citet{NitschkeVoigt_AiDE_2025}, 
an additional constraint force $ \FbND = \fbND + \fnorND\normal $ may need to be accounted in the fluid equations \eqref{eq:model_conforming_tangentialeq} and \eqref{eq:model_conforming_normalseq}.
Spatial state variation yields
\begin{align*}
	 \innerH{\tangentR}{\FbND, \Wb} 
    	&:= -\innerH{\tangentR}{\deltafrac{\energyND}{\para}, \Wb} 
    	  = \int_{\surf} \eth_{\Wb}\inner{\tangentQS}{\lambdabND,\qb} + \inner{\tangentQS}{\lambdabND,\qb}(\DivC\Wb) \dS\\
    	  &= \innerH{\tangentQR}{ \eth_{\Wb}\lambdabND + (\DivC\Wb)\lambdabND, \qb}
    	     + \innerH{\tangentQR}{\lambdabND, (\eth_{\Wb}\circ\projQS\circ\projS[^2])\Qb}
\end{align*}
for all virtual displace displacements $\Wb\in\tangentR $ and $\qb=(\projQS\circ\projS[^2])\Qb$.
The first summand vanishes due to constraint \eqref{eq:ND_constraint}.
For the second summand we use the identity \citep{NitschkeSadikVoigt_IJoAM_2023}
\begin{align*}
    \eth_{\Wb}\IdS
    	&= \eth_{\Wb}\left( \Id - \normal\otimes\normal \right)
    	 = -\eth_{\Wb}\normal\otimes\normal - \normal\otimes\eth_{\Wb}\normal
    	 = \normal(\nablaC\Wb)\otimes\normal + \normal\otimes\normal\nablaC\Wb \formComma
\end{align*}
and obtain by means of \eqref{eq:ND_constraint} that
\begin{align*}
    \left(\projQS\circ\projS[^2]\circ\eth_{\Wb}\circ\projQS\circ\projS[^2]\right)\Qb\hspace{-8em}\\
    	&= \left(\projQS\circ\projS[^2]\right) \Big(\IdS\eth_{\Wb}\Qb\IdS + \eth_{\Wb}\IdS\Qb\IdS + \IdS\Qb\eth_{\Wb}\IdS \\
    	&\hspace{8em} - \frac{\eth_{\Wb}\Qb\dbdot\IdS + \Qb\dbdot\eth_{\Wb}\IdS}{2}\IdS - \frac{\Qb\dbdot\IdS}{2}\eth_{\Wb}\IdS\Big)\\
    	&= \left(\projQS\circ\projS[^2]\right) \Big(\IdS\eth_{\Wb}\Qb\IdS + \normal\otimes\normal(\nablaC\Wb)\Qb\IdS + \IdS\Qb(\normal(\nablaC\Wb))\otimes\normal\\
    	&\hspace{8em} - \frac{\eth_{\Wb}\Qb\dbdot\IdS}{2}\IdS - \frac{\Qb\dbdot\IdS}{2}\left( \normal(\nablaC\Wb)\otimes\normal + \normal\otimes\normal\nablaC\Wb \right)\Big) \\
    	&= \left(\projQS\circ\projS[^2]\right)\Big( \IdS\eth_{\Wb}\Qb\IdS - \frac{\eth_{\Wb}\Qb\dbdot\IdS}{2}\IdS \Big)
    	 = \left(\projQS\circ\projS[^2]\right)\eth_{\Wb}\Qb
\end{align*}
holds, since, a priori, $\Qb$ is surface conforming and thus cannot have tangential-normal, \resp\ normal-tangential, components.
Eventually, the spatial state variation results in
\begin{align*}
    \innerH{\tangentR}{\FbND, \Wb}
    	&= \innerH{\tangentQR}{\lambdabND, \eth_{\Wb}\Qb}
    	 = \innerH{\tangentQR}{\HbND, \eth_{\Wb}\Qb}\formPeriod
\end{align*}
This gives a pure ``gauge of surface independence'' force \wrt\ $\Qb$, \cf\ \citet{NitschkeSadikVoigt_IJoAM_2023}.
Such gauge forces can be ignored, since the net gauge force is balanced within the fluid equation, independently of the choice of $ \eth_{\Wb}\Qb $, see \citet{NitschkeVoigt_AiDE_2025}.
Therefore, we choose $ \eth_{\Wb}\Qb = \nullb $ \oeda, and obtain $\FbND= \nullb$, \ie\ no additional constraint forces have to be to consider in \eqref{eq:model_conforming_tangentialeq} and \eqref{eq:model_conforming_normalseq}.
To summarize the implications: 
In order to apply the ``No flat Degeneracy'' constraint \eqref{eq:ND_constraint} to the surface conforming model \eqref{eq:model_conforming} without using the Lagrange multiplier $\lambdabND$, 
it suffices to omit the flat degenerated molecular equation \eqref{eq:model_conforming_qtensor} and substitute $\qb=\nullb$ throughout the remaining model.

\section{Numerical realization}\label{sec:Numerics}
The numerical approach to solve the surface Beris--Edwards model for symmetric bilayers \eqref{eq:BE_R3} and the hydrodynamic surface Landau--Helfrich model for asymmetric bilayers \eqref{eq:full_R3}, extends the approach used in \cite{KrauseVoigt_JoCP_2023} for \eqref{eq:H_R3} and \citep{BachiniKrauseEtAl_JoFM_2023,SischkaEtAl_CMAME_2025} for surface two-phase systems. We consider a surface finite element method (SFEM) \citep{Dziuk_2013,nestler2019finite} within an Arbitrary Lagrangian-Eulerian (ALE) approach \citep{KrauseVoigt_JoCP_2023,Sauer_JoFM_2025}. This includes a mesh redistribution approach \citep{Barrett_SIAMJSC_2008} for which the systems are extended by equations for the surface parametrization
\begin{align}
	\partial_t \para \cdot \normal &= \Vb\cdot\normal \label{eq:normalvel}\\
	\meanc \normal &= \Delta_C \para \,, \label{eq:meandiff}
\end{align}
which generate a tangential mesh movement to maintain shape regularity and additionally provide an implicit representation of the mean curvature $\meanc$.

The implementation is done in DUNE/AMDiS \citep{vey2007amdis,witkowski2015software,Dune2.10} using the DUNECurvedGrid library \citep{dune-curvedGrid}. We consider a discrete $k$-th order approximation $\surf_h^k$ of $\surf$, with $h$ the size of the mesh elements, i.e. the longest edge of the mesh, and consider each geometrical quantity like the normal vector $\normal_h$, the identity $\IdS{}_{,h}$, the shape operator $\shop_h$, and the Gaussian curvature $\gaussc_h$ 
with respect to $\surf_h^k$. 
We define the discrete function spaces for scalar functions by $\discspace(\surf_h^k)=\{ \psi \in C^0(\surf_h^k) \vert \psi\vert_{T}\in\mathcal{P}_{k}(T)\}$ and for vector fields by $\discvecspace(\surf_h^k)=[\discspace(\surf_h^k)]^3$. Within these definitions $T$ is the mesh element and $\mathcal{P}_{k}$ are the polynomials of order $k$. We consider  $\Vb_h,\para\in\discvecspace[2](\surf_h^k)$,  $\meanc_h, \beta_h \in \discspace[2](\surf_h^k)$, and $p_h\in \discspace[1](\surf_h^k)$, which leads to an isogeometric setting and a $\mathcal{P}_{2}-\mathcal{P}_{1}$ Taylor-Hood element for the surface Navier--Stokes-like equations. We discretize in time using semi-implicit time stepping with constant step size $\tau$. We use an operator splitting approach. In each time step we first solve the equation for the orientation field and then solve the surface Navier--Stokes-like equations and the mesh redistribution together. Then, we update the surface, lifting the solutions $\Vb_h^n$, $\tp_h^n$, $\meanc_h^n$ and $\beta_h^n$ to the new surface and computing the remaining geometric quantities $\normal_h^n$, $\IdS{}_{,h}$ $\shop_h^n$ and $\gaussc_h^n$ for the new surface. To resolve the non-linear terms in the dynamic of the orientation field, we consider all surface quantities from the previous timestep and consider a one-step Picard iteration for the terms non-linear in $\beta$. For the surface dynamic, we consider $\normal_h$, $\IdS{}_{,h}$, $\shop_h$ and $\gaussc_h$ explicitly and resolve the non-linearity in the material acceleration $(\nablaC \Vb_h) \Vb_h$ by considering $\nablaC \Vb_h$ implicitly and $\Vb_h$ explicitly. For convergence studies of the surface Navier--Stokes equations with this approach we refer to \citep{KrauseVoigt_JoCP_2023}, for such studies with similar coupling terms with surface scalar quantities we refer to \citep{BachiniKrauseEtAl_JoFM_2023} and for corresponding treatments of Gaussian curvature terms see \citep{SischkaEtAl_CMAME_2025,SischkaNitschkeEtAl_FD_2025}.

In order to explore the differences between the surface Beris--Edwards model and the hydrodynamic surface Landau--Helfrich model with the corresponding surface Navier--Stokes--Helfrich models, we consider an initial surface $\surf_0$ as a perturbed unit sphere
\begin{align}
	\surf_0 = \left\{1+r_0Y^m_l(\phi,\vartheta): \phi \in [0,\pi],\vartheta\in[-\pi,\pi] \right\}, \quad r_0>0, \\
	Y^m_l(\phi,\vartheta) = \sqrt{\frac{2l+1(l-m)!}{4\pi(l+m)!}}P^m_l(\cos\vartheta)e^{im\phi}
\end{align}
with spherical harmonics $Y^m_l$ and Legendre polynomials $P^m_l$ and the case $l = 5$, $m = 3$ and $r_0 = 0.5$. The velocity field is initialized with $\Vb_0 = 0$, the scalar order field with $\beta_0 = \frac{2}{3}$. The initial surface has non-zero mean and Gaussian curvature and is out of equilibrium. 

As physical parameters, we choose $\xi = 0$, $\coeffIF = M = \rho = 1$ and $L = 0.1$. In case of the surface Beris--Edwards model \eqref{eq:BE_R3} we further consider $a=0$, $b=-27$ and $c=\frac{27}{2}$, in case of the hydrodynamic surface Landau--Helfrich model \eqref{eq:full_R3}, we choose $\hat{a} = 0$, $\varpi = 1$ and $\kappa=\bar\kappa=0.2$ leading to comparable conditions (c.f. \eqref{eq:consistency}). In the hydrodynamic surface Landau--Helfrich model, we further choose $\meanc_0 = -1.77$, which corresponds to the mean curvature of the expected equilibrium shape of $\surf$ for both systems, a sphere with radius $r=1.129$. We further consider $\hat\meanc_0 = 0$. Numerical parameters, such as mesh size $h$ and time step $\tau$ are chosen to resolve the highest curvature values and to meet stability constraints, following \citep{KrauseVoigt_JoCP_2023, BachiniKrauseEtAl_JoFM_2023, SischkaEtAl_CMAME_2025}. To study the influence of the orientation field $\beta$ on the system, we also solve the surface Navier--Stokes--Helfrich model \eqref{eq:H_R3} with either $H_0 = 0$ for the symmetric case or $H_0 = -1.77$ for the asymmetric case and all other parameters chosen as stated above.

\begin{description}
\item[Funding:] 
    A.V. was supported by German Research
    Foundation (DFG) through FOR3013 "Vector- and tensor-valued surface PDEs".
\item[Competing interests:] 
    There are no competing interests.
\item[Authors' contributions:] 
    This project was conceived by I.N., J.M.S. and A.V.; I.N. derived the theory, the results were analyzed by I.N. and A.V., J.M.S. performed the simulations and the main text was written by I.N. and A.V..
\end{description}

%
%

\clearpage
\printbibliography

\end{document}

%% file: symbols.tex
\newcommand{\wrt}{w.\,r.\,t.}
\newcommand{\eg}{e.\,g.}
\newcommand{\soth}{s.\,t.} 
\newcommand{\ie}{i.\,e.}

\newcommand{\oeda}{w.\,l.\,o.\,g.}

\newcommand{\resp}{resp.}
\newcommand{\cf}{cf.}

\newcommand{\formComma}{\,\text{,}}
\newcommand{\formPeriod}{\,\text{.}}

\newcommand{\R}{\mathbb{R}}

\newcommand{\surf}{\mathcal{S}}

\newcommand{\Vcal}{\mathcal{V}} 

\newcommand{\tangent}[2][]{\tensor{\operatorname{T}\!}{#1}#2}
\newcommand{\tangentS}[1][]{\tangent[#1]{\surf}}
\newcommand{\tangentR}[1][]{\tangent[#1]{\R^3\vert_{\surf}}}
\newcommand{\tangentQS}{\tensor{\operatorname{\mathcal{Q}}\!}{^2}\surf}
\newcommand{\tangentQR}{\tensor{\operatorname{\mathcal{Q}}\!}{^2}\R^3\vert_{\surf}}

\newcommand{\tangentScal}{\tangentS[^0]}

\newcommand{\dbdot}{\operatorname{:}}

\newcommand{\timeD}[1][\Phi]{\operatorname{d}_{t}^{#1}\!}
\newcommand{\timeJ}{\mathfrak{J}}

\newcommand{\Dt}[1][]{\operatorname{D}^{#1}_t \!}
\newcommand{\Dmat}{\Dt[\mfrak]}

\newcommand{\Dlow}{\Dt[\flat]}
\newcommand{\Djau}{\Dt[\jau]}

\newcommand{\Comp}{\mathsf{C}}
\newcommand{\nablaC}{\nabla_{\!\Comp}}

\newcommand{\Tr}{\operatorname{Tr}}
\renewcommand{\div}{\operatorname{div}}
\newcommand{\Div}{\operatorname{Div}}

\newcommand{\DivC}{\Div_{\!\Comp}}

\newcommand{\Grad}{\operatorname{Grad}}
\newcommand{\GradC}{\Grad_{\Comp}}

\newcommand{\hil}{\operatorname{L}^{\!2}}
\newcommand{\hilspace}[1]{\hil(#1)}
\newcommand{\inner}[3][0]{\left\langle #3 \right\rangle_{\hspace{-#1pt}#2}}
\newcommand{\normsq}[3][0]{\left\| #3 \right\|_{\hspace{-#1pt}#2}^2}
\newcommand{\innerH}[3][0]{\inner[#1]{\hilspace{#2}}{#3}}
\newcommand{\normHsq}[3][0]{\normsq[#1]{\hilspace{#2}}{#3}}

\newcommand{\proj}{\operatorname{\Pi}}
\newcommand{\projS}[1][]{\proj_{\tangentS[#1]}}
\newcommand{\projQS}{\proj_{\tangentQS}}

\newcommand{\paraC}{X}
\newcommand{\para}{\boldsymbol{\paraC}}
\newcommand{\normalC}{\nu}
\newcommand{\normal}{\boldsymbol{\normalC}}
\newcommand{\shopC}{I\!I}
\newcommand{\shop}{\boldsymbol{\shopC}}
\newcommand{\meanc}{\mathcal{H}}
\newcommand{\gaussc}{\mathcal{K}}

\newcommand{\Id}{\boldsymbol{Id}}
\newcommand{\IdS}{\Id_{\surf}}

\newcommand{\mfrak}{\mathfrak{m}}
\newcommand{\ofrak}{\mathfrak{o}}
\newcommand{\jau}{\mathcal{J}}

\newcommand{\ab}{\boldsymbol{a}}

\newcommand{\eb}{\boldsymbol{e}}
\newcommand{\fb}{\boldsymbol{f}}

\newcommand{\hb}{\boldsymbol{h}}

\newcommand{\qb}{\boldsymbol{q}}
\newcommand{\rb}{\boldsymbol{r}}

\newcommand{\vb}{\boldsymbol{v}}
\newcommand{\wb}{\boldsymbol{w}}

\newcommand{\Ab}{\boldsymbol{A}}

\newcommand{\Cb}{\boldsymbol{C}}

\newcommand{\Eb}{\boldsymbol{E}}
\newcommand{\Fb}{\boldsymbol{F}}
\newcommand{\Gb}{\boldsymbol{G}}
\newcommand{\Hb}{\boldsymbol{H}}
\newcommand{\Ib}{\boldsymbol{I}}

\newcommand{\Qb}{\boldsymbol{Q}}
\newcommand{\Rb}{\boldsymbol{R}}
\newcommand{\Sb}{\boldsymbol{S}}

\newcommand{\Vb}{\boldsymbol{V}}
\newcommand{\Wb}{\boldsymbol{W}}

\newcommand{\zetab}{\boldsymbol{\zeta}}
\newcommand{\etab}{\boldsymbol{\eta}}
\newcommand{\thetab}{\boldsymbol{\theta}}      

\newcommand{\lambdab}{\boldsymbol{\lambda}}

\newcommand{\sigmab}{\boldsymbol{\sigma}}      

\newcommand{\Sigmab}{\boldsymbol{\Sigma}}

\newcommand{\Psib}{\boldsymbol{\Psi}}

\newcommand{\tp}{\tilde{p}}
\newcommand{\tsigmab}{\tilde{\sigmab}}
\newcommand{\tSigmab}{\tilde{\Sigmab}}

\newcommand{\bp}{\bar{p}}

\newcommand{\vnor}{v_{\bot}}
\newcommand{\anor}{a_{\bot}}
\newcommand{\fnor}[1][]{f^{\bot}_{#1}}
\newcommand{\nullb}{\boldsymbol{0}}

\newcommand{\potenergy}{\mathfrak{U}}
\newcommand{\energy}{\mathfrak{E}}
\newcommand{\fluxpotential}{\mathfrak{R}}
\newcommand{\power}{\mathfrak{P}}

\newcommand{\EL}{\textup{EL}}
\newcommand{\energyEL}{\potenergy_{\EL}} 

\newcommand{\hbEL}{\hb_{\EL}}
\newcommand{\zetabEL}{\zetab_{\EL}}
\newcommand{\omegaEL}{\omega_{\EL}}

\newcommand{\sigmabEL}{\sigmab_{\EL}}

\newcommand{\el}{\textup{el}}
\newcommand{\energytel}{\tilde{\potenergy}_{\el}} 
\newcommand{\energyel}{\potenergy_{\el}} 

\newcommand{\THT}{\textup{TH}} 
\newcommand{\energyTH}{\potenergy_{\THT}} 

\newcommand{\hbTH}{\hb_{\THT}}
\newcommand{\omegaTH}{\omega_{\THT}}

\newcommand{\BE}{\textup{BE}}

\newcommand{\fnorBE}{\fnor[\BE]}

\newcommand{\coeffIF}{\upsilon}

\newcommand{\IM}{\textup{IM}}
\newcommand{\energyIM}[1][\Phi]{\fluxpotential_{\IM}^{#1}}

\newcommand{\zetabIM}[1][\Phi]{\zetab_{\IM}^{#1}}
\newcommand{\omegaIM}[1][\Phi]{\omega_{\IM}^{#1}}

\newcommand{\sigmabIM}[1][\Phi]{\sigmab_{\IM}^{#1}}

\newcommand{\AC}{\textup{AC}}
\newcommand{\energyAC}{\fluxpotential_{\AC}}

\newcommand{\sigmabAC}{\sigmab_{\AC}}
\newcommand{\FbAC}{\Fb_{\AC}}

\newcommand{\IA}{\textup{IA}}
\newcommand{\NA}{\textup{NA}}

\newcommand{\cNA}{\alpha_{\NA}}

\newcommand{\sigmabNA}{\sigmab_{\NA}}

\newcommand{\NV}{\textup{NV}}
\newcommand{\energyNV}{\fluxpotential_{\NV}} 

\newcommand{\hbNV}{\hb_{\NV}}
\newcommand{\omegaNV}{\omega_{\NV}}

\newcommand{\sigmabNV}{\sigmab_{\NV}}
\newcommand{\tsigmabNV}{\tilde{\sigmab}_{\NV}}

\newcommand{\SC}{\textup{SC}}

\newcommand{\ND}{\textup{ND}}
\newcommand{\energyND}{\potenergy_{\ND}}
\newcommand{\lambdabND}{\lambdab_{\ND}} 
\newcommand{\HbND}{\Hb_{\ND}}
\newcommand{\hbND}{\hb_{\ND}}
\newcommand{\FbND}{\Fb_{\ND}}
\newcommand{\fbND}{\fb_{\ND}}
\newcommand{\fnorND}{\fnor[\ND]}

\newcommand{\LdG}{\textup{LdG}}
\newcommand{\LH}{\textup{LH}}
\newcommand{\energyLdG}{\potenergy_{\LdG}}
\newcommand{\energyLH}{\potenergy_{\LH}}

\newcommand{\TOT}{\textup{TOT}}

\newcommand{\Cset}{\mathcal{C}} 

\newcommand{\deltafrac}[2]{\frac{\delta #1}{\delta #2}}
\newcommand{\ddfrac}[2][]{\frac{\textup{d} #1}{\textup{d} #2}}
\newcommand{\ddt}[1][]{\ddfrac[#1]{t}}

\newcommand{\dS}{\textup{d}\surf}

\newcommand{\discspace}[1][k]{W_{#1}}
\newcommand{\discvecspace}[1][k]{\boldsymbol{W}_{#1}}